\documentclass{aa}  
\usepackage{natbib}
\bibpunct{(}{)}{;}{a}{}{,} 
\usepackage{graphicx}
\usepackage{txfonts}
\usepackage{tikz}
\usepackage{bbm}
\usepackage{hyperref}
\usepackage[normalem]{ulem}
\usepackage{color}
\definecolor{darkraspberry}{rgb}{0.53, 0.15, 0.34}
\definecolor{color1}{rgb}{0.3, 0.7, 0}

\begin{document}


  \title{Simulation-based inference has its own Dodelson-Schneider effect (but it knows that it does)}
   

   \author{
        J. Homer,
        \inst{1, 2}
        \fnmsep\thanks{email: jed.homer@physik.lmu.de}
        O. Friedrich,
        \inst{1, 2, 3}
        \and
        D. Gruen
        \inst{1, 2, 3}
    }

   \institute{
   University Observatory, Faculty of Physics, Ludwig-Maximilians-Universität, Scheinerstr. 1, 81677 Munich, Germany. 
   \and
   Munich Center for Machine Learning (MCML).
   \and 
   Excellence Cluster ORIGINS, Boltzmannstr. 2, 85748 Garching, Deutschland.  
   }
   
    

   \date{Received X; accepted X}


  \abstract
   {Making inferences about physical properties of the Universe requires knowledge of the data likelihood. A Gaussian distribution is commonly assumed for the uncertainties with a covariance matrix estimated from a set of simulations. The noise in such covariance estimates causes two problems: it distorts the width of the parameter contours, and it adds scatter to the location of those contours which is not captured by the widths themselves. For non-Gaussian likelihoods, an approximation may be derived via Simulation-Based Inference (SBI). It is often implicitly assumed that parameter constraints from SBI analyses, which do not use covariance matrices, are not affected by the same problems as parameter estimation with a covariance matrix estimated from simulations.}
   {We measure the coverage and marginal variances of the posteriors derived using density estimation SBI, over many identical experiments, to investigate whether SBI suffers from effects similar to those of covariance estimation in Gaussian likelihoods.
   }
   {We use Neural Posterior and Likelihood Estimation with continuous and masked autoregressive normalizing flows for density estimation. We fit our approximate posterior models to simulations drawn from a Gaussian linear model, so that the SBI result can be compared to the true posterior and effects related to noise in the covariance estimate are known analytically. We test linear and neural network based compression, demonstrating that neither methods circumvent the issues of covariance estimation.}
   {SBI suffers an inflation of posterior variance that is equal or greater than the analytical result in covariance estimation for Gaussian likelihoods for the same number of simulations. This inflation of variance is captured conservatively by the reported confidence intervals, leading to an acceptable coverage regardless of the number of simulations. The assumption that SBI requires a smaller number of simulations than covariance estimation for a Gaussian likelihood analysis is inaccurate. The limitations of traditional likelihood analysis with simulation-based covariance remain for SBI with finite simulation budget. Despite these issues, we show that SBI correctly draws the true posterior contour given enough simulations.}
   {}


\titlerunning{SBI has its own Dodelson-Schneider effect (but it knows that it does)}
\authorrunning{J. Homer, O. Friedrich and D. Gruen}
\maketitle

%

\section{Introduction}

Current- and next-generation cosmological experiments such as the Dark Energy Survey (DES, \citealt{des}), Euclid \citep{Euclid}, the Nancy Grace Roman Space Telescope \citep{Roman}, the Vera C. Rubin Observatory Legacy Survey of Space and Time (LSST, \citealt{LSST}), the 4m Multi-Object Spectroscopic Telescope (4MOST, \citealt{4MOST}) and the Dark Energy Spectroscopic Instrument (DESI, \citealt{DESI}) will return a huge volume of observational data of the large scale-structure in the Universe. The purpose of this effort is to constrain the values of fundamental physical parameters as accurately and precisely as possible. The limit of the cosmological information that can be extracted from a measurement ultimately depends on the typically unknown likelihood function of the data. The likelihood function compares the data to a theoretical model and this comparison may be inaccurate and imprecise because 
\begin{itemize}
    \item the model prediction for the expectation value as a function of the parameters may not be known analytically or it may be inaccurately predicted in numerical simulations, 
    \item in addition to the expectation value, the likelihood function, which gives the distribution of the measurement around the expectation, is not known.
\end{itemize} 

The likelihood functions for observables of the large-scale structure are often unknown and only approximate expressions for them exist \citep{cora1pt}. This is particularly the case for higher-order statistics. These obtain information beyond the amount extracted by traditional two-point functions \citep{beyond2pt} - and can potentially break parameter degeneracies - with many approaches based on measuring higher order correlations of cosmological fields having been proposed \citep{Kacprzak2016, cora1pt, dssdaniel, dssoliver, Hamaus2020, Contarini2023, anik, waveletST, chrispeaks, SIMBIGHou}. This promise but lack of analytical expressions for their likelihood promotes the use of SBI (or `likelihood-free', \citealt{SBI}) methods to extract information from such statistics. 

Simulation-based inference (SBI, \citealt{SBI}) covers a broad class of statistical techniques (e.g. \citealt{ETH_ABC, NRE, papamakarios1, lueckmann, AlsingDELFI, TMNRE22}) that derive an approximate likelihood or posterior model from a set of simulations paired with their model parameters.  The likelihood is fit with no assumptions on the data-generating process and allows for complex effects in the measurement process to be forward-modelled. This is in contrast to classical or explicit likelihood methods that require an analytic model for the expectation value and statistical uncertainties in the data.  An additional claimed benefit of SBI methods is the ease of using multiple probes together without analytically modelling cross-correlations between the measurements \citep{10x2pt, Reeves12x2pt}. The density estimation techniques used for SBI \citep{AlsingDELFI, delfi_nuisance, lueckmann, papamakarios1, variationalinferencesbi} apply generative models fit with either maximum-likelihood optimisation or variational inference.

  \begin{figure*}
    \centering
       \includegraphics[width=18cm]{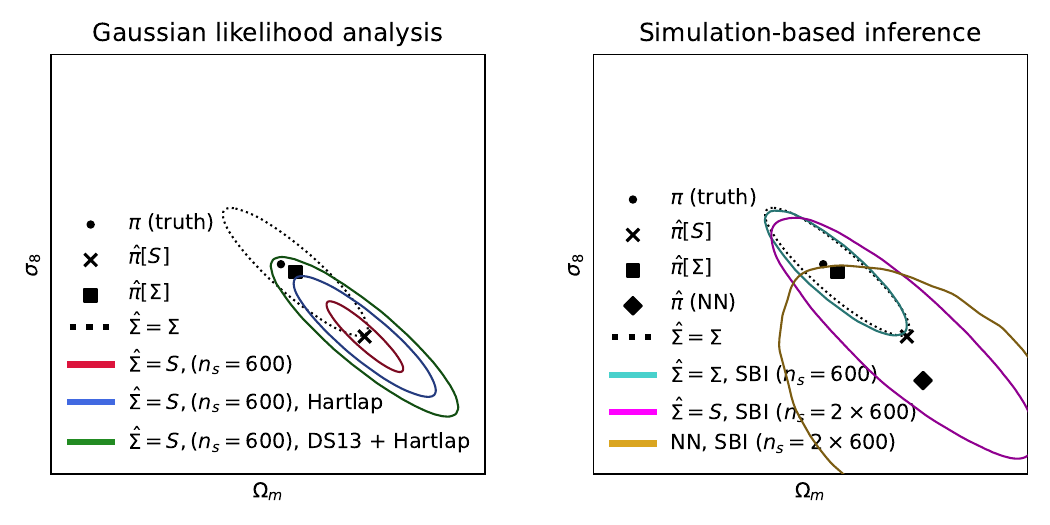}
       \caption{Effects of an estimated data error distribution on the posterior contours and their locations. We show estimates $\hat{\boldsymbol{\pi}}$ of parameters $\boldsymbol{\pi}$ and the estimated $1-\sigma$ confidence contours. The black dotted contours show the posterior derived with the true covariance $\Sigma$ from noiseless data. \textit{Left}: The results from a Gaussian likelihood analysis. In red, the posterior derived with the estimated covariance $S$ from 600 simulations. In blue, the posterior derived the same way, but with the inverse covariance corrected with the Hartlap factor \citep{Anderson2003, Hartlap2006}. Finally, in green, the same but with the correction of \cite{Dodelson2013} applied to the inverted sample covariance as well, arguably the best result obtainable without additional knowledge about the data covariance. We ask, which of the contours in this diagram will SBI methods produce given a similarly limited number of simulations? \textit{Right}: The results of applying SBI with linear and neural network based compressions. In magenta, the contour returned from applying Neural Likelihood Estimation with $600$ simulations compressed using a data covariance $S$ estimated with $600$ simulations. In teal, Neural Likelihood estimation applied with $600$ simulations compressed using the true data covariance $\Sigma$. In yellow, the contour obtained by applying Neural Likelihood estimation applied with $600$ simulations compressed using a neural network that was fit with $600$ simulations.}
       \label{fig:cov_example2}
    \end{figure*} 

The use of SBI is now established in many branches of cosmology with a variety of different methods for deriving posteriors from sets of simulations; via density-estimation of the likelihood or posterior \citep{AlsingDELFI, ETH_ABC, Makinen_2021, LeclerqHeavens, BOLFI}, ABC \citep{ETH_ABC, ABC_SNIa}, Neural Ratio Estimation \citep{NRE, TMNRE22}, which has been applied to galaxy clustering \citep{Beatriz, Modi_galaxy_clustering, SIMBIG, SIMBIG_2, SIMBIGCNN, SIMBIGHou}, weak-lensing \citep{LFI_weaklensing, JeffreySVSBI, DESGattiSBI, JeffreyDESSBI}, cosmic-shear \citep{LinShear, SBIKiloDegree}, cluster abundance \citep{Amara}, cosmic microwave background radiation \citep{TMNRE22}, gravitational wave sirens \citep{Gerardi}, type Ia supernovae \citep{ABC_SNIa} and the cosmic 21cm signal \citep{21cmSBI}, emerging as a fast and efficient method for deriving posteriors from measurements after fitting to forward-modelled simulations of data. The potential of SBI methods is claimed to allow for Bayesian inference with high-dimensional data with unknown or inaccurate models for the expectation value and likelihood that are difficult or intractable to analyze using traditional likelihood-based methods.



With the issues discussed for modern observational cosmology, the potential of SBI methods for these problems and the rapid innovations within the machine learning literature to implement the analyses, we ask

\begin{center}
    \emph{Can SBI methods return posterior estimates whose locations scatter less compared to the true posterior than those of a Gaussian likelihood with simulation-estimated covariance, for the same number of simulations?}
\end{center}
and in doing so, 
\begin{center}
    \emph{Does SBI know about the additional scatter in location of its posterior estimates? In other words, does SBI inflate its contours sufficiently (compared to the true posterior) to still achieve good parameter coverage in repeated experiments?} 
\end{center}

To answer these questions we test the SBI framework for Gaussian data vectors with a parameter independent covariance matrix and linear parameter dependence for the expectation value of the data. This allows us to compare our SBI posterior estimates, in which the likelihood or data covariance is not known, to both the true posterior and to the posteriors that would be derived from covariance estimation within a Gaussian likelihood assumption. As a data vector we assume a tomographic cosmic shear data vector, and we test different techniques to compress this data: score compression \citep{score_compression} with and without knowledge of the data vectors covariance as well as a neural network based compression. We then investigate how SBI combined with either of these compression techniques performs with respect to the the scatter, width and coverage of resulting parameter contours as a function of the number of simulations used for training (again, compared to the true posterior and posteriors derived from covariance estimation). These experiments test the common assertion that SBI is \textit{not} affected by the errors that make covariance estimation impracticable for high-dimensional data vectors and computationally expensive simulations \citep[e.g.][]{JeffreySVSBI, JeffreyDESSBI, DESGattiSBI}. In Figure \ref{fig:cov_example2} we illustrate our questions with a set of posteriors derived with a Gaussian likelihood analysis - accounting for the unknown covariance - contrasted with SBI analyses with either compression method.

Our data are generated from a Gaussian likelihood with a model that is linear in its parameters, in which case a Fisher analysis, the noise bias correction of \citet{Hartlap2006}, and the derivation of the excess scatter in \citet{Dodelson2013} are all valid. This allows us to analytically determine the best-fit parameters and the confidence contours in each likelihood analysis for comparison with the posteriors derived with SBI.

In Section \ref{section:covariance} we review the main issues for estimating data covariances and methods for accounting for the noise in covariances estimated from simulations, highlighting where SBI is claimed to be advantageous. In Section \ref{section:methods} we outline the methods for density estimation, SBI, data compression used in this work and a simple experiment in which we test these methods \footnote{Our code is a available at \href{https://github.com/homerjed/sbiax}{github.com/homerjed/sbiax}}. In Section \ref{section:results} we present results from the experiments and in Section \ref{section:conclusions} we conclude. 

    \begin{figure*}
    \centering
       \includegraphics[width=12.0cm]{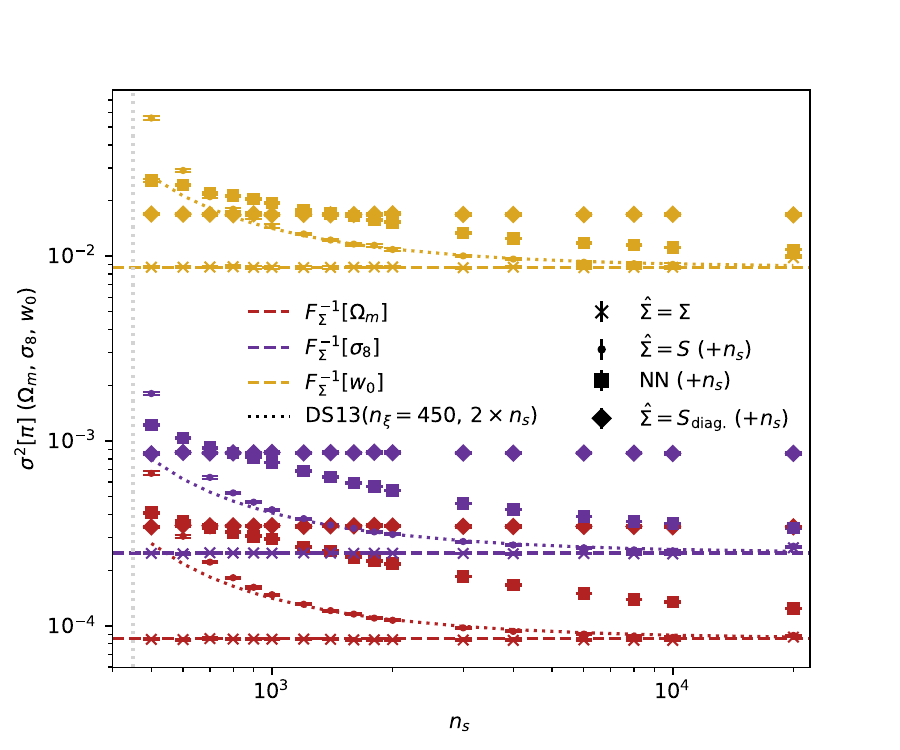}
       \caption{Average model parameter posterior variance, reported by the methods compared in this work, conditioned on noisy datavectors estimated with Neural Likelihood estimation using a masked autoregressive flow. The colour-coded dashed lines show the (Fisher) variances that would have been measured if the data covariance was known exactly and used in a Gaussian likelihood ansatz with a flat prior. Cross points label posterior variances from SBI analyses where the data covariance was known exactly for linear compression (Equation \ref{eq:freq_map}). Dotted lines show the expected variances of the MAPs that would have been measured when using a data covariance estimated from a set of $n_s$ simulations. Note that these lines multiply the Fisher variance with the factor (Equation \ref{eq:f_ds}) by \cite{Dodelson2013} calculated using $2\times n_s$ simulations. Circle points label posterior variances from SBI analyses where the data covariance was estimated from $n_s$ simulations and used in a linear compression. Diamond points label posterior variances obtained when using a compression with the diagonal elements of the estimated covariance matrix. Square points show variances obtained from using a neural network for the compression, trained on $n_s$ simulations. The additional simulations, not labeled on the $x$-axis, but required for the separate compressions (where the true covariance is unknown) are noted for each method. When the true data covariance is not known, requiring the use of double the number of simulations, the reconstructed posterior errors from SBI are significantly higher than the \cite{Dodelson2013} corrected errors when that correction is substantial.}
       \label{fig:widths_nle_maf}
    
    \centering
       \includegraphics[width=18cm]{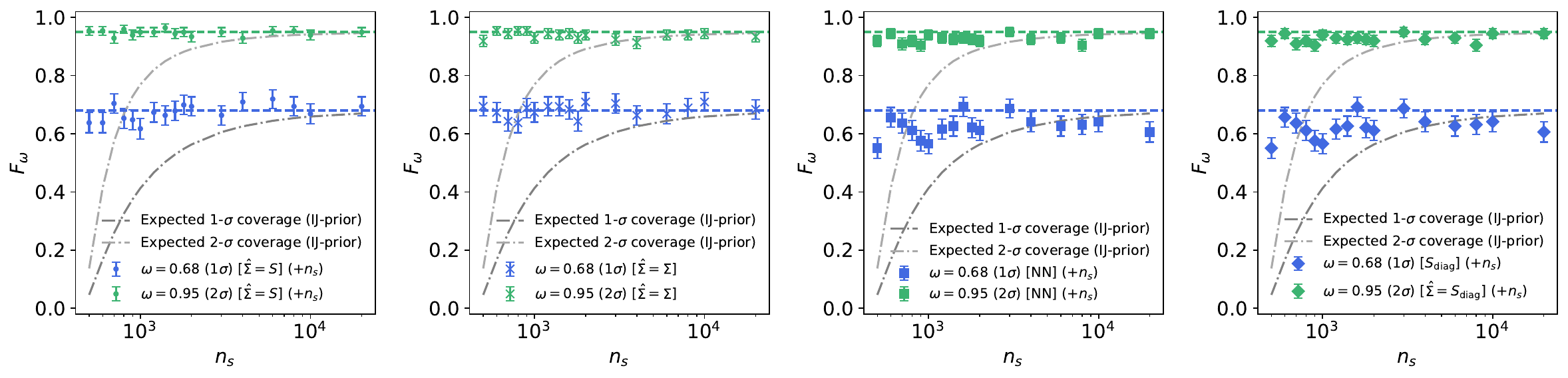}
       \caption{Coverage $F_\omega$, i.e. how often the true cosmology in the experiment is found inside the 68\% ($1-\sigma$) and 95\% ($2-\sigma)$) credible regions of the estimated posterior (see Equation~ \ref{eq:expected_coverage}) against the number of simulations $n_s$ used for the training set. Shown is the result for Neural Likelihood Estimation (with a MAF model) for independently sampled data vectors and data covariance matrices in a series of repeated experiments with the number of simulations for each experiment on the horizontal axis. The expected coverage of a Gaussian posterior with a debiased estimate of the precision matrix (using the Hartlap correction, Equation \ref{eq:HartlapFactor}) and posterior covariance corrected with the \cite{Dodelson2013} factor is plotted for both coverage intervals with dashed lines. The grey-toned lines show the expected coverage of the common approach using a Gaussian posterior with a precision matrix corrected by applying the Hartlap factor. The additional simulations, not labeled on the $x$-axis, but required for the separate compressions (where the true covariance is unknown) are noted for each method. The SBI posteriors obtain the correct coverage to within errors for all numbers of simulations $n_s$ and each compression method.}
       \label{fig:coverage_nle_maf}
    \end{figure*} 

\section{Parameter estimation with noisy covariance matrices}\label{section:covariance}

We study a simple inference problem in which the likelihood is not known. A common ansatz in cosmological analyses is to assume a Gaussian likelihood for a given set of statistics of the large-scale structure. Typically this statistical model for the likelihood depends on a parameterised expectation value $\boldsymbol{\xi}[\boldsymbol{\pi}]$ and a covariance matrix $\Sigma[\boldsymbol{\pi}]$. If the true covariance is not known, then an estimate $\hat{\Sigma}$ may be derived from simulated data. Often, this estimator is derived for one fiducial set of parameters, and the parameter-dependence of the covariance is assumed to be negligible. Note however, that this is not in principle a limitation of covariance estimation within the Gaussian likelihood assumption. It is possible to derive estimates $\hat\Sigma[\boldsymbol{\pi}]$ at different values of the parameters $\boldsymbol{\pi}$ (and e.g.\ interpolate between them). We emphasize this, because the parameter-dependence of the (width of the) likelihood function is not the main reason why SBI may be preferable compared to covariance estimation paired with a Gaussian likelihood assumption. Instead, SBI is most needed in cases where the Gaussian assumption itself (whether with or without parameter-dependent covariance) is insufficient. 

The Gaussian likelihood assumption can be justified when the data are made up of measurements in independent sub-volumes of a survey area so that their sum is Gaussian distributed via the central limit theorem. In practice, the validation of this ansatz is demanding \citep{KIDScovariance, DEScovariance} and can resort in scale-cuts to the datavector, reducing the information return of the analysis. 

However, even when the Gaussian likelihood ansatz is justified, {covariance estimation only yields noisy versions of the true covariance $\Sigma$ and its inverse $\Sigma^{-1}$ (the \textit{precision} matrix). If independent and accurately drawn realisations $\boldsymbol{\xi}_i$ ($i = 1\ ,\ \dots\ ,\ n_s$) of the data vector in simulated data are available, then an unbiased estimate for the data covariance $\Sigma$ is calculated as
\begin{equation}\label{eq:S}
    S = \frac{1}{n_s - 1}\sum_{i=1}^{n_s} (\boldsymbol{\xi}_i - \boldsymbol{\bar{\xi}}) (\boldsymbol{\xi}_i - \boldsymbol{\bar{\xi}})^T,
\end{equation}
where
\begin{equation}
    \boldsymbol{\bar{\xi}} = \frac{1}{n_s}\sum_{i=1}^{n_s} \boldsymbol{\xi}_i \ ,
\end{equation}
\noindent is the mean calculated over all available data. The inversion of this matrix is a non-linear process which causes $S^{-1}$ to be a biased estimator of $\Sigma^{-1}$ and to have significantly poorer noise properties than $S$ \citep{Taylor2012}.

If the true likelihood function is indeed Gaussian, then the bias of $S^{-1}$ can be corrected by applying a factor $h$ as \citep{Kaufman, Hartlap2006} 
\begin{equation} \label{eq:HartlapFactor}
    \hat{\Sigma}^{-1} = (hS)^{-1},
\end{equation}
where 
\begin{equation}
    h = \frac{n_s - 1}{n_s - n_{\boldsymbol{\xi}} - 2}.
\end{equation}
The above correction only de-biases $\hat{\Sigma}^{-1}$ and not the full likelihood function which even in the Gaussian case is a non-linear function of $\hat{\Sigma}^{-1}$. Surprisingly though, at least the width of the likelihood function and of the parameter posteriors that might be derived from it are usually quite well approximated by just inserting $(hS)^{-1}$ into the Gaussian likelihood function \citep{PME, Percival2021}. The main problem of covariance estimation is not the width of parameter posteriors but the location of the latter. The noise in $(hS)^{-1}$ causes additional scatter in the location of parameter contours which can make the overall scatter of maximum-posterior parameters in repeated experiments much larger than indicated by the posterior width itself \citep{Dodelson2013, PME}.

It is this effect of increased parameter scatter that we are after! Naive covariance estimation together with the correction factor $h$ does not by itself understand that it suffers from this scatter (e.g. Figure 1 of \citealt{PME}). This raises the question: why should SBI - in which we do not just estimate the covariance but the full likelihood shape from simulations - not suffer from this effect? And if it does suffer from it, does SBI at least automatically adjust its own contour size to account for this additional scatter in maximum-posterior parameters?

In the case of covariance estimation, it was derived by \cite{Dodelson2013} that the scatter of maximum-posterior parameters (or rather: their parameter covariance) is enhanced compared to inverse Fisher matrix $F_{\Sigma}^{-1}$ (i.e.\ compared to the naive expectation) via
\begin{equation}\label{eq:dodelson}
    \langle (\boldsymbol{\hat{\pi}} - \boldsymbol{\pi})(\boldsymbol{\hat{\pi}} - \boldsymbol{\pi})^T \rangle_{\boldsymbol{\xi}, S} = [1 + B(n_{\boldsymbol{\xi}} - n_{\boldsymbol{\pi}})]F_{\Sigma}^{-1},
\end{equation}
\noindent where $n_{\boldsymbol{\xi}}$ is the number of data points, $n_{\boldsymbol{\pi}}$ is the number of parameters and the factor $B$ is given by 
\begin{equation}\label{eq:B}
    B = \frac{n_s - n_{\boldsymbol{\xi}} - 2}{(n_s - n_{\boldsymbol{\xi}} - 1)(n_s - n_{\boldsymbol{\xi}} - 4)}\ .
\end{equation}
In order for parameter contours to take into account this additional uncertainty, the covariance of parameter posteriors should be enhanced by the Dodelson-Schneider factor
\begin{align}\label{eq:f_ds}
    f_{\mathrm{DS}} =&\ 1 + \dfrac{(n_{\boldsymbol{\xi}} - n_{\boldsymbol{\pi}})(n_s - n_{\boldsymbol{\xi}} - 2)}{(n_s - n_{\boldsymbol{\xi}} - 1)(n_s - n_{\boldsymbol{\xi}} - 4)} \\
    \approx&\ 1 + \dfrac{n_{\boldsymbol{\xi}} - n_{\boldsymbol{\pi}}}{n_s - n_{\boldsymbol{\xi}}}
\end{align}
where the approximation in the last line is valid if $n_s \gg n_{\boldsymbol{\xi}} \gg n_{\boldsymbol{\pi}}\,$. A concrete recipe for how to widen parameter posterior in order to correct for the Dodelson-Schneider effect (and other, usually subdominant effects) has been provided by \cite{Percival2021}. There, the authors marginalise over the unknown, true covariance $\Sigma$ in a Bayesian approach that takes into account the likelihood function $p(S|\Sigma)$, given by the Wishart distribution in the Gaussian case, and a prior distribution $p(\Sigma) \propto |\Sigma|^m\,$. They provide the coefficient $m$ in the prior distribution that leads to the desired frequentist coverage of the resulting Bayesian parameter constraints. (Note especially, that the coefficient $m$ considered by \cite{SellentinHeavens2015} does not lead to that desired coverage.)

As mentioned before, we want to investigate whether SBI suffers from problems analogous to the ones described above. And if it does - whether it self corrects to obtain the desired coverage properties or whether a manual widening of posteriors as in the case of Gaussian covariance estimation is required. Note however: even if SBI automatically corrects for its own Dodelson-Schneider effect, this correction still means that posteriors are widened and this parameter information is diluted. This may significantly hinder SBI approaches to deliver on the promised improvements of parameter constraints compared to likelihood-full analyses of summary statistics whose uncertainties can be modelled analytically.

One more comment to emphasize that the Dodelson-Schneider effect can not be easily circumvented: it is not possible to beat down $f_{\mathrm{DS}}$ by simply compressing a given set of statistics down do a smaller data vector. E.g.\ for the popular MOPED compression \citep{MOPED} (or equivalently score compression \citep{score_compression}), this would require knowledge of the covariance $\Sigma$ in the first place. And if that is approximated by an estimate $S$, then this is just shifting the problem from one side of statistical analysis to another. In fact, \cite{Dodelson2013} (herafter DS13) derived $f_{\mathrm{DS}}$ exactly by considering the scatter of optimally compressed statistics that use noisy covariances.

Given the described assumptions, and upon acquiring a measurement $\boldsymbol{\hat{\xi}}$, inference on the values of model parameters is based on a Gaussian likelihood and the estimated precision matrix $\hat{\Sigma}^{-1}$ 

\begin{equation}
    p(\boldsymbol{\hat{\xi}}|\boldsymbol{\pi}, \hat{\Sigma}^{-1}) \propto \exp \bigg [ -\frac{1}{2}\chi^2 (\boldsymbol{\hat{\xi}}, \boldsymbol{\pi}, \hat{\Sigma}^{-1}) \bigg],
\end{equation}

where 

\begin{equation}
   \chi^2(\boldsymbol{\hat{\xi}}, \boldsymbol{\pi}, \hat{\Sigma}^{-1}) \equiv (\boldsymbol{\hat{\xi}} - \boldsymbol{\xi}[\boldsymbol{\pi}]) \hat{\Sigma}^{-1} (\boldsymbol{\hat{\xi}} - \boldsymbol{\xi}[\boldsymbol{\pi}])^T.
\end{equation}

A posterior distribution for the parameters in light of the measurement, using a prior density $p(\boldsymbol{\pi})$, is expressed as 

\begin{equation}\label{eq:bayes}
    p(\boldsymbol{\pi}|\boldsymbol{\hat{\xi}}) \propto p(\boldsymbol{\hat{\xi}}|\boldsymbol{\pi}, \hat{\Sigma}^{-1})p(\boldsymbol{\pi}),
\end{equation}

\noindent ignoring a parameter independent normalisation factor.

For a given $S$ and $\boldsymbol{\hat{\xi}}$, the \textit{maximum a posteriori} (MAP) or \textit{best-fit} parameter estimates $\boldsymbol{\hat{\pi}}$ are obtained with an estimated precision matrix $\hat{\Sigma}^{-1}=(hS)^{-1}=\Sigma^{-1} + \Delta_{\Sigma^{-1}}$ as

\begin{equation}\label{eq:freq_map}
    \boldsymbol{\hat{\pi}} = \boldsymbol{\pi} + [F_\Sigma + \Delta_F]^{-1}\partial_{\boldsymbol{\pi}}\boldsymbol{\xi}[\boldsymbol{\pi}]^T [\Sigma^{-1} + \Delta_{\Sigma^{-1}}] (\boldsymbol{\hat{\xi}} - \boldsymbol{\bar{\xi}}).
\end{equation}

Here $F_\Sigma$ is the Fisher matrix, a function of the likelihood, which is written as 

\begin{equation}\label{eq:fisher}
    F_{\Sigma}(\boldsymbol{\pi})=\partial_{\boldsymbol{\pi}}\boldsymbol{\xi}[\boldsymbol{\pi}]^T \Sigma^{-1} \partial_{\boldsymbol{\pi}}\boldsymbol{\xi}[\boldsymbol{\pi}],
\end{equation}

for a Gaussian likelihood parameterised with the true precision matrix $\Sigma^{-1}$. The quantity $\Delta_F$ is of the same form with the error on the precision matrix $\Delta_{\Sigma^{-1}}$ instead of the unknown true precision matrix $\Sigma^{-1}$. For a model $\boldsymbol{\xi}[\boldsymbol{\pi}]$ - linear in $\boldsymbol{\pi}$ with Gaussian error bars on the data - the Fisher matrix defined above exactly quantifies the information content of the data upon the model at a fixed point in parameter space. 

With the increase in the dimensionality of measurements returned from cutting-edge cosmological surveys, it may not be possible to obtain a precision matrix due to the singularity of the estimated covariance matrix \citep{Hartlap2006, Dodelson2013, WhitePadmanabhanCov}. In particular this is true for datavectors that combine multiple probes which are critical to break parameter degeneracies and calibrate systematic effects independently \citep{fluri2, 10x2pt, Reeves12x2pt}. In this case, estimating one from simulations requires an unfeasible amount of computation \citep{taylor2013, PME}. 

However, if it is possible, a covariance matrix $\Sigma$ may be estimated from data realisations themselves (e.g. Jacknifing or sub-sampling \citep{norbergjackknife, oliverjackknife, percivaljacknife}, a set of accurate numerical simulations that assumed an underlying cosmological model \citep{Percival2014, cora1pt}, a theoretical covariance model \citep{Schneider2002, DEScovariance, KIDScovariance, Planckcovariance, cosmolike, ShearcovarianceLinke, 10x2pt, Reeves12x2pt} or some hybrid method combining these techniques \citep{PME, Hallhybridcovariance}. Alternatively, covariance matrices computed from an analytical covariance model are noiseless and easily invertible, but are only accurate given a good understanding of the statistical properties of the data. Ultimately, obtaining an analytic covariance may be an insurmountable task.
    
\section{Methods}\label{section:methods}

In the following subsections we describe the data likelihood we run our analyses with, the experiments with SBI, the data compression, the normalizing flow models we use for density estimation and hyperparameter optimisation.

\subsection{Model}

We run an inference of cosmological parameters estimated from a measurement, drawn from a linearised model of the DES-Y3 cosmic shear 2-point function data vector, where noise is sampled from a fiducial covariance matrix. This data covariance is calculated with an analytic halo-based model from \cite{Krause2017}. The linear model is a Taylor expansion of the full model around the fiducial point in parameter space $\boldsymbol{\pi}^0$, written as

\begin{equation}
    \boldsymbol{\xi}[\boldsymbol{\pi}] = \boldsymbol{\xi}[\boldsymbol{\pi}^0] + (\boldsymbol{\pi} - \boldsymbol{\pi}^0)^T \partial_{\boldsymbol{\pi}} \boldsymbol{\xi}[\boldsymbol{\pi}] |_{\boldsymbol{\pi}=\boldsymbol{\pi}^0}.
\end{equation}
   
The true data likelihood from which we sample data in our experiments is written 
$\mathcal{G}[\boldsymbol{\hat{\xi}}|\boldsymbol{\xi}[\boldsymbol{\pi}], \mathbf{\Sigma}]$. We use a uniform prior on the parameters where $\Omega_m\in[0.05, 0.55]$, $\sigma_8\in[0.45, 1.0]$ and $w_0\in[-1.4, -0.1]$. The parameter set that maximises the likelihood (and therefore the posterior in this case) is given by Equation \ref{eq:freq_map}. 

The datavector consists of two-point functions of cosmic shear measurements \citep{Schneider2002, Schneider_2006, Kilbinger_2015, Amon_shear_2022}. The shear field is a map of coherent distortions from weak gravitational lensing of galaxy images in the large-scale structure matter distribution. The measurement is sensitive to the density fluctuations projected along the line of sight and weighted by the lens galaxy distribution. It is known to measure a degenerate combination of $\sigma_8$ and the matter density parameter $\Omega_m$. 
  
\subsection{Experimental setup}

We run a simple inference on a set of parameters from a noisy datavector with a linear compression. Note that since we assume linear expectation value and constant covariance, this is equivalent to score compression \citep{score_compression, delfi_nuisance}. This experiment is repeated for when the data covariance is known and separately when it is estimated from a set of $n_s$ simulations. When the data covariance is estimated, the linear compression increases the scatter in the estimated parameters. This affects the compression on the noisy datavector as well as the simulated data that is used to fit the normalizing flow density estimators. We also run separate SBI experiments that use a neural network for the compression - this would be assumed to be a simple fix to the issues of covariance matrix estimation.

We measure the marginal uncertainty on the inferred parameters by calculating the variance of samples from the posteriors estimated with SBI conditioned on the noisy datavectors. We also measure the coverage of the posteriors for each datavector. Each analysis consists of separately applying Neural Posterior Estimation (NPE, \citealt{npe}) and Neural Likelihood Estimation (NLE, \citealt{nle}) to fit a posterior given a set of simulations and their model parameters. See Table \ref{table:experiments} for a description of the separate analyses we run. The total number of simulations available in the experiment mimics the number of available $N$-body simulations in a cosmological analysis. We repeat the NPE and NLE analyses 200 times for experiments where the data covariance is known and another 200 times for where it is estimated from simulations. 

In one experiment we 
    \begin{itemize}
        \item sample a set of `true' parameters from a uniform prior $\boldsymbol{\pi}\sim p(\boldsymbol{\pi})$,
    
        \item initialise the density estimator parameters randomly, 
        
        \item generate a set of $n_s$ simulations 
        \begin{itemize}
            \item at the fiducial parameters $\boldsymbol{\pi}$ sampling $\{\boldsymbol{\xi}_i\}_{i=1}^{n_s} \sim \mathcal{G}[\boldsymbol{\xi}|\boldsymbol{\xi}[\boldsymbol{\pi}], \Sigma]$, and calculate the sample covariance matrix $S$ if $\Sigma$ is unknown, \textit{or}
            \item at a set of parameters sampled from the prior $p(\boldsymbol{\pi})$ to train a neural network to compress our data,
        \end{itemize}
        
        \item sample physics parameters from a uniform prior $\{\boldsymbol{\pi}_i\}_{i=1}^{n_s} \sim p(\boldsymbol{\pi})$,
        
        \item generate $n_s$ simulations from the $n_s$ prior samples, using the linearised model with noise sampled from the true covariance matrix $\Sigma$, to compress the simulations with
        
        \begin{itemize}
            \item a linear compression $\boldsymbol{\hat{\pi}}=\boldsymbol{\pi} + F^{-1}_{\hat{\Sigma}^{-1}} \partial_{\boldsymbol{\pi}} \boldsymbol{\xi}[\boldsymbol{\pi}]\hat{\Sigma}^{-1}(\boldsymbol{\hat{\xi}} - \bar{\boldsymbol{\xi}})$, parameterised by true expectation $\xi[\boldsymbol{\pi}]$, the true or estimated precision ($\Sigma^{-1}$ or $(hS)^{-1}$), and the theory derivatives $\partial_{\boldsymbol{\pi}}\boldsymbol{\xi}[\boldsymbol{\pi}]$, \textit{or} 
            \item a neural network trained on the first set of $n_s$ simulations and parameter pairs,
        \end{itemize}

        \item fit a normalizing flow to the set of compressed simulations and parameters $\{ \hat{\boldsymbol{\xi}}_i, \boldsymbol{\pi}_i\}$ by maximising the log-probability (of the likelihood or posterior) with stochastic gradient descent,

        \item sample the posterior given a measurement $\boldsymbol{\hat{\xi}}$, compressed to $\hat{\boldsymbol{\pi}}$, from the true data-generating likelihood.
    \end{itemize}
This experiment is idealised in the following ways
    \begin{itemize}
        \item the measurement errors upon our data $\boldsymbol{\hat{\xi}}$ are drawn from the same distribution from which the simulations used to fit the density estimation models are drawn from,
        \item there are no nuisance parameters in our modelling of the data to marginalize over,
        \item the analytic compression of our data is (given enough simulations at the fiducial parameters) lossless, so that the posterior given the summary is identical to the posterior given the data,
        \item the true expectation value $\bar{\boldsymbol{\xi}}=\langle\hat{\boldsymbol{\xi}}\rangle$ lies in the parameter space meaning there are parameters $\boldsymbol{\pi}$ such that $\boldsymbol{\xi}[\boldsymbol{\pi}] = \bar{\boldsymbol{\xi}}$,
        \item the Dodelson-Schneider correction factor (Equation \ref{eq:dodelson}) is exact given that the Fisher information of a Gaussian likelihood (with a model that is linear in the parameters) quantifies the posterior covariance.
    \end{itemize}

Note that NLE requries a specific prior in order to calculate the posterior whereas NPE implicitly uses the prior defined by the distribution of parameters that are used to generate the training data. We therefore adopt the priors used to sample the parameters (in the NLE analyses) for generating the training data of the flows to ensure both methods use the same prior to allow for a comparison of equivalent analyses. 

The number of simulations $n_s$ input to an experiment using either NPE or NLE depends on the compression method being used. This is shown in Table \ref{table:experiments} for reference. To compress our data we use either a linear compression with $\hat{\Sigma}=\Sigma$, $\hat{\Sigma}=S$ or $\hat{\Sigma}=S_{\text{diag.}}$ and, separately, a neural network fit with a mean-squared error loss. Either option, except when $\Sigma$ is known, requires $2n_s$ simulations in total - $n_s$ for the covariance estimation or neural network training and $n_s$ for fitting the density estimator of the likelihood or posterior. The comparison of the posteriors from SBI therefore should be compared, for our Gaussian linear model, with a Gaussian likelihood analysis using $2n_s$ simulations.

SBI methods fit the model $\boldsymbol{\xi}[\boldsymbol{\pi}]$, covariance $\Sigma$ and likelihood shape simultaneously from simulations. In Appendix \ref{appendix:fitting_expectation} we use the same Gaussian linear model to test if fitting the expectation alone, with a known covariance, introduces significant uncertainty in the posterior as a function of $n_s$. Since the sample mean and covariance are independent, this would only affect analyses for which $n_{\boldsymbol{\xi}} \sim n_{\boldsymbol{\xi}}^2$, which is a concern for future experiments. However, this shows that the tests presented in this work are fair for SBI - the fitting of the model alongside the covariance has almost no effect - even for the lowest number of simulations we consider in our experiments.

\subsection{Compressing the data}

For density estimation it is advantageous to reduce the dimensionality of the data. In the case of a linear compression \citep{Tegmark, MOPED, score_compression} one implicitly assumes a model to derive the statistics \citep{HeavensExtreme} and the sampling distribution of the statistics is Gaussian only if the data errors are Gaussian. Neural network based summary statistics \citep{fluri1, fluri2, imnn, imnn21cm, Villanueva_Domingo_2022} can easily be fit to data (which are typically non-standard summary statistics), though they have no analytic likelihood for the summary given the input and so extracting credible intervals is not currently possible, except from the use SBI. Additionally, there is no guarantee that fitting a neural network to regress the model parameters of input data will produce an unbiased estimator of the parameters, since the MSE estimator is only the maximum-likelihood estimator for Gaussian distributed data with unit covariance \citep{MurphyML}. 

In the experiments in which $\hat{\Sigma}=\Sigma$ or $\hat{\Sigma}=S$ the data are linearly compressed via Equation \ref{eq:freq_map} to $n_{\boldsymbol{\pi}}$ summaries so that the normalizing flow likelihood model is fit to a Gaussian likelihood with a Fisher matrix given by either $F_{\Sigma}$ or $F_S$ depending on whether the covariance is known or not. The noisy data covariance (as a function of $n_s$) in our experiments limits the amount of information the normalizing flow posterior or likelihoods can extract about the model parameters. In the case that the precision matrix is known exactly and the model $\boldsymbol{\xi}[\boldsymbol{\pi}]$ is linear in the parameters, the linear compression conserves the information content of the data $\boldsymbol{\hat{\xi}}$. 

We also test the use of a neural network $f_\psi$ in compressing the data. The neural network consists of simple linear layers and non-linear activations. A simulation is input to the network and the parameters of the network $\psi$ are obtained by stochastic gradient descent of the mean-squared error loss

\begin{equation}\label{eq:mse}
    \Lambda(\boldsymbol{\xi}, \boldsymbol{\pi};\psi) = ||f_\psi(\boldsymbol{\xi}) - \boldsymbol{\pi}||^2_2\ .
\end{equation}

\subsection{Density estimation with normalizing flows}

\begin{table}
\begin{center}
\begin{tabular}{|c|ccccc|}
 \hline
     & $\hat{\Sigma}=\Sigma$ & $\hat{\Sigma}=S$ & $\hat{\Sigma}=S_{\text{diag.}}$ & NN & \\ [0.5ex] 
 \hline 
 NPE &  C, M  & C, M & C, M & C, M & \\ 
     &  $(n_s)$ & $(2n_s)$ & $(2n_s)$ & $(2n_s)$ & \\ 
 \hline
 NLE &  C, M & C, M & C, M & C, M & \\ 
     &  $(n_s)$ & $(2n_s)$ & $(2n_s)$ & $(2n_s)$ & \\ 
 \hline
\end{tabular}
\caption{Describing the experiments we run. For each method of density estimation SBI (NPE or NLE) a density estimator model (either a CNF, denoted `C' or MAF, denoted `M') is fit to simulations compressed with $\hat{\Sigma}$ or a neural network (NN). Shown also are multiples of the number of simulations $n_s$ input to each experiment, depending on the compression methods. We ran 200 experiments with independent datavectors and covariance matrices (when estimated) for each combination in this table.}
\label{table:experiments}
\end{center}
\end{table}

In order to derive posteriors from a measurement using density estimation SBI methods, a density estimator is fit to pairs of model parameters and simulated data to estimate the likelihood or posterior directly \citep{SBI, lueckmann, papamakarios1, AlsingDELFI}. normalizing Flows \citep{originalflows0, originalflows1, rezendeflows, flows} are a class of generative models that fit a sequence of bijective transformations from a simple base distribution to a complex data distribution. The transformation is estimated directly from the simulation and parameter pairs via minimising the KL-divergence between the unknown likelihood (posterior) and the flow likelihood (posterior). We use Masked Autoregressive Flows (MAFs, \citealt{mafs}) and Continuous Normalizing Flows (CNFs, \citealt{ffjord, neuralode}). We use CNFs in order to adopt some of the latest density estimation techniques from the machine learning literature.\footnote{Note that diffusion \citep{ddpm, Songsbgm} and flow-matching models \citep{cfm} calculate log-likelihoods by approximating continuous flows.} Two different models are used to validate and compare the performance of the density estimation by either. 

Normalizing flows transform data $\boldsymbol{x}$ to Gaussian distributed samples $\boldsymbol{y}$. This mapping, when conditioned on physics parameters $\boldsymbol{\pi}$ and parameters $\phi$ of a neural network $f_{\phi}$, is written as

\begin{equation}\label{eq:flow_map}
    \boldsymbol{y} = f_\phi(\boldsymbol{x};\boldsymbol{\pi}),
\end{equation}

where $\boldsymbol{y} \sim \mathcal{G}[\boldsymbol{y}|\mathbf{0}, \mathbb{I}]$ and $\mathbb{I}$ is the identity matrix. An exact log-likelihood estimate of the probability of a datapoint conditioned on physics parameters can be calculated with a normalizing flow by using a change-of-variables between $\boldsymbol{y}$ and $\boldsymbol{x}$ expressed as 

\begin{equation}
    \log p_\phi(\boldsymbol{x}|\boldsymbol{\pi}) = \log \mathcal{G}[f_\phi(\boldsymbol{x};\boldsymbol{\pi})|\boldsymbol{0}, \mathbb{I}] + \log \big | \mathbf{J}_{f_\phi}(\boldsymbol{x};\boldsymbol{\pi})\big |,
\end{equation}

where $\mathbf{J}_{f_\phi}$ is the Jacobian of the normalizing flow transform $f_\phi$ in Equation \ref{eq:flow_map}.

These bijective transformations between two densities can be composed to produce more complex distributions by using separate transformations in a sequence. For a normalizing flow with $K$ transforms $\{g_k\}_{k=0}^{K}$ in sequence parameterised as $\phi=\{\phi_k\}_{k=0}^K$, the log-likelihood of the flow is written

\begin{align} \label{eq:flow}
    \log p_\phi(\boldsymbol{x}|\boldsymbol{\pi}) 
    &= \log \mathcal{G}([g_{\phi_K}\circ g_{\phi_{K-1}} \circ ... \circ g_{\phi_0}](\boldsymbol{x};\boldsymbol{\pi})|\boldsymbol{0}, \mathbb{I}) \nonumber \\
    &+ \; \sum_{k=0}^K \log \big | \mathbf{J}_{g_{\phi_k}}(\boldsymbol{x};\boldsymbol{\pi})\big | \ .
\end{align}

We note that while it is common to use ensembles of density estimators together for diagnosing the fit of the likelihood models \citep{AlsingDELFI, JeffreySVSBI, DESGattiSBI, JeffreyDESSBI}, this should not be necessary for the simple Gaussian linear model we use here.

To obtain an approximate likelihood $p_\phi(\boldsymbol{\pi}|\boldsymbol{x})$ (for NLE) or posterior model $p_\phi(\boldsymbol{x}|\boldsymbol{\pi})$ (for NPE) we fit a normalizing flow to a set of simulations and parameters. The parameters of the normalizing flow model $\phi$ that maximise the log-likelihood $p_\phi(\boldsymbol{x}|\boldsymbol{\pi})$ (or log-posterior) are obtained by minimising the forward KL-divergence between the unknown likelihood (posterior) $q(\boldsymbol{x}|\boldsymbol{\pi})$ and the normalizing flow likelihood (posterior) $p_\phi(\boldsymbol{x}|\boldsymbol{\pi})$.

The loss function for the normalizing flow is then given by 

\begin{align}\label{eq:KL}
    \langle D_{KL}(q||p_\phi) \rangle_{\boldsymbol{\pi}} &= \int \text{d}\boldsymbol{\pi} \; p(\boldsymbol{\pi}) \int \text{d}\boldsymbol{x} \; q(\boldsymbol{x}|\boldsymbol{\pi}) \log \frac{q(\boldsymbol{x}|\boldsymbol{\pi})}{p_\phi(\boldsymbol{x}|\boldsymbol{\pi})}, \nonumber \\
    &= \int \text{d}\boldsymbol{\pi} \int \text{d}\boldsymbol{x} \; p(\boldsymbol{\pi}, \boldsymbol{x})[\log q(\boldsymbol{x}|\boldsymbol{\pi}) - \log p_\phi(\boldsymbol{x}|\boldsymbol{\pi})], \nonumber \\
    &\geq -\int \text{d}\boldsymbol{\pi} \int \text{d}\boldsymbol{x} \; p(\boldsymbol{x},\boldsymbol{\pi}) \log p_\phi(\boldsymbol{x}|\boldsymbol{\pi}), \nonumber \\
    &\approx -\frac{1}{N}\sum_i^N \log p_\phi(\boldsymbol{x}_i|\boldsymbol{\pi}_i)\ .
\end{align}

Note that terms independent of $\phi$ are dropped since their derivative with respect to $\phi$ is zero. This implies the loss function for the normalizing flow is given by 

    \begin{equation}
        \Lambda(\{\boldsymbol{x},\boldsymbol{\pi}\};\phi) = -\frac{1}{N}\sum_{i=1}^N \log p_\phi(\boldsymbol{x}_i|\boldsymbol{\pi}_i),
    \end{equation}

for NLE and 

    \begin{equation}
        \Lambda(\{\boldsymbol{x},\boldsymbol{\pi}\};\phi) = -\frac{1}{N}\sum_{i=1}^N \log p_\phi(\boldsymbol{\pi}_i|\boldsymbol{x}_i),
    \end{equation}

for NPE (since the derivation of Equation \ref{eq:KL} applies in the same way to the posterior). The CNF and MAF models we use for our experiments are described in detail in Appendix \ref{appendix:flows}.

\subsection{Architecture and fitting}
Our CNF model has 1 hidden layer of 8 hidden units with $\tanh(\cdot)$ activation functions and an ODE solver timestep of $0.1$. We train using early-stopping with a patience value of 40 epochs \citep{doubledescent}. The MAF models had 5 MADE transforms, parameterised by neural networks with 2 layers and 32 hidden units using $\tanh(\cdot)$ activation functions. The MAFs used a patience of 50 epochs. We use the ADAM \citep{adam} optimiser with a learning rate of $10^{-3}$ for stochastic gradient descent of the negative-log likelihood loss with both density estimator models.

\subsection{Hyperparameter optimisation}
It is not computationally feasible in SBI analyses to optimise simultaneously for the architecture and parameterisation of a density estimation model when fitting the likelihood or posterior. Since the reconstructed posterior depends on these hyperparameters implicitly we must run our experiment separately to obtain the best settings when tested on data that is not part of the training set. 

The parameters of the architecture and optimisation procedure are tuned by experimentation using \texttt{optuna} \citep{optuna} to find the parameters that minimise an optimisation function over repeated experiments where the true data covariance is known exactly. The experiment uses a further $10^4$ data samples for training. The average log-likelihood (or log-posterior) of the flow was calculated on a separate test set of $10^4$ simulation and parameter pairs. The architecture and training parameters that maximised this average log-likelihood were chosen. See Appendix \ref{appendix:hyperparameters} for details of the hyperparameter optimisation.

\section{Results}\label{section:results}

In the next section we discuss the quantitative results of measuring the posterior widths (Section \ref{section:results_widths}) and the coverage over repeated experiments (Section \ref{section:results_coverages}).
   
\subsection{Posterior widths}\label{section:results_widths}

Figure \ref{fig:widths_nle_maf} shows the reconstructed posterior widths $\sigma^2[\boldsymbol{\pi}]$ from repeated identical experiments as a function of the number of simulations $n_s$ used to fit the normalizing flows (and where appropriate - estimating the data covariance or training a neural network for compression - which requires another $n_s$ simulations). The results for the other experiments (noted in Table \ref{table:experiments}) are shown in Appendix \ref{appendix:extra_results}.

The DS13 factor (Equation \ref{eq:f_ds}) sets the expected width of the scatter in the parameter estimators when both the model $\boldsymbol{\xi}[\boldsymbol{\pi}]$ and covariance $\Sigma$ are determined by data. The factor is calculated for $2n_s$ simulations (the size of the training set plus those used to estimate the data covariance that is used for the compression), with the number of bins in the data equal to the uncompressed data dimension of $n_{\boldsymbol{\xi}}=450$ bins. For each value of $n_s$ the factor is plotted to compare the expected posterior width in a Gaussian likelihood analysis - where the covariance is estimated from a set of simulations - to the SBI analyses. By default, every SBI analysis uses $n_s$ simulations to train the normalising flows.

The SBI analyses plotted for this comparison are repeated with four compressions. The first is a linear compression (Equation \ref{eq:freq_map}) in which the true covariance is known, e.g. $\hat{\Sigma}=\Sigma$. The second is a linear compression with the simulation-estimated covariance $\hat{\Sigma}=S$ (calculated with $n_s$ simulations). Third is a linear compression with only the diagonal elements of this same covariance $\hat{\Sigma}=S_{\text{diag.}}$. Last is a compression parameterised with a neural network that is trained on an additional $n_s$ simulations. 

For the experiments where the true covariance is known, SBI can obtain posterior widths equal to the Fisher errors for all $n_s$ values as expected. The posterior widths for the $\hat{\Sigma}=S$ compression converge closely to the DS13 factor at around $2\times 10^4$ simulations (for the training set of the flows plus the covariance estimation) which means the flow models have fit the correct likelihood shape. The $\hat{\Sigma}=S_{\text{diag.}}$ experiments (plotted with diamonds) return posterior widths that are less optimal, over all values of $n_s$, compared to the marginal Fisher variances and the SBI analyses that use $\hat{\Sigma}=\Sigma$. Despite the posterior width being significantly increased, for low values of $n_s$ of around a few hundred simulations, the posterior widths obtained when using $\hat{\Sigma}=S_{\text{diag.}}$ are below the Fisher variances expanded by the DS13 factor - i.e. the posterior width when estimating the full covariance - for the same $n_s$ input to the experiment. This is simply due to the fact that the inversion of a diagonal matrix $S_{\text{diag.}}$ does not combine the noisy off-diagonal elements - of the matrix $S$ that is inverted - in a non-linear way. None the less, the structure of the estimated covariance is incorrect, which inflates the variance of the summaries. This increases the marginal variances of the posteriors obtained by the flows. 

The experiments involving a neural network for the compression function show a significantly different relation of the posterior width to the number of simulations compared to the $\hat{\Sigma}=\Sigma$ and $\hat{\Sigma}=S$ experiments. The widths from the neural network compression experiments are not the same as any of the widths using linear compression since the network does not invert the true (or estimated) data covariance. Rather, if the network calculates a compression close to an optimal linear compression - it is possible that the network down-weights the noisier datavector elements, to minimise the MSE loss, for a lower number of simulations.  This is not the same as the inverse-variance weighting of the linear compression in Equation \ref{eq:freq_map}. Curiously, this effect - within the regime of $n_s \leq n_{\boldsymbol{\xi}}$ for a given value of $n_{\boldsymbol{\pi}}$ - allows one to obtain an average posterior width that is \textit{smaller} than that of a posterior using an estimated covariance adjusted with the DS13 factor (calculated with $2n_s$ simulations) despite the fact that the covariance is not known. How optimal the summary by a neural network is depends strongly on the training hyperparameters, optimisation method and choice of loss function - though quantifying the information content is only possible with an analytic posterior such as in this work.

\subsection{Coverage}\label{section:results_coverages}

The expected coverage probability of the SBI posterior estimators measures the proportion of repeated identical experiments where a credible region of the posterior estimator contains the true parameter set used to generate the datavector. The coverage probability quantifies how conservative or overconfident the posterior estimator is compared to the true posterior. 

We define $F_\omega$ as the fraction of experiments where the true cosmology $\boldsymbol{\pi}$ is inside the $68.3\%$ ($\omega=0.68$) or $95.4\%$ ($\omega=0.95$) confidence contour around the MAP $\boldsymbol{\hat{\pi}}$. The fraction of $J$ posterior samples with posterior probability under the SBI estimator less than that of the true data-generating parameters $\boldsymbol{\pi}_i$ for the $i$-th experiment is the empirical coverage probability of the $i$-th posterior, written as

\begin{equation}
    f(\boldsymbol{x}_i, \boldsymbol{\pi}_i, \phi_i) = \frac{1}{J} \sum_{j=1}^J \mathbbm{1}[p_{\phi_i}(\boldsymbol{\pi}_i|\boldsymbol{x}_i) > p_{\phi_j}(\boldsymbol{\pi}_j|\boldsymbol{x}_i)],
\end{equation}

where $\mathbbm{1}[\cdot]$ is the indicator function and $\boldsymbol{\pi}_j$ is the $j$-th posterior sample from the $i$-th posterior. The fraction of experiments that obtain a coverage probability $\omega$ is calculated as

\begin{equation}\label{eq:expected_coverage}
    F_\omega = \frac{1}{n_e} \sum_{i=1}^{n_e} \mathbbm{1} [ f(\boldsymbol{x}_i, \boldsymbol{\pi}_i, \phi_i) > \omega ],
\end{equation}

over a set of independent but identical experiments with independently sampled data $\boldsymbol{\hat{\xi}}$, data covariance matrices $S$ and density estimator model parameters $\phi_i$. If the true covariance is known, then $F_\omega$ should be equal to $0.68$ for the $1-\sigma$ region and $0.95$ for the $2-\sigma$ region if the data are sampled from the true likelihood and the posterior estimator has converged. 

For the NLE experiments using an MAF Figure \ref{fig:coverage_nle_maf} (see Appendix \ref{appendix:extra_results} for the remaining experiments in Table \ref{table:experiments}) shows the coverages measured over the repeated experiments. Either density estimation method obtains the correct coverage to within errors, regardless of the model used and of whether the data covariance is known or not. This suggests that the normalizing flow likelihoods correctly expand their contours to account for the MAP scatter when using an estimated covariance or a trained neural network for compression which would be expected given the hierarchical modelling by the normalising flow of the functional form of the likelihood (including the expectation and covariance) to calculate the likelihoods. 

It should be noted that, as found in \cite{PME} and \cite{Percival2021} for the same $n_{\boldsymbol{\xi}}$ and $n_{\boldsymbol{\pi}}$ in our experiments, the use of the independence Jeffreys prior on true covariance matrix in \cite{SellentinHeavens2015} gives a posterior with a model parameter covariance that matches a Gaussian posterior after scaling the data covariance matrix by the Hartlap factor and applying the factor in \cite{Dodelson2013} (Equation \ref{eq:dodelson}) however the posterior does not take into account the Dodelson-Schneider factor. Hence in Figures \ref{fig:coverage_nle_maf} (see also Appendix \ref{appendix:extra_results}) we plot the analytic expected coverage for the Gaussian posterior with a scaled parameter variance and debiased precision matrix, accounting for the simulations used to estimate the covariance as well as the training-set simulations. The SBI methods both are able to obtain the correct coverage for all $n_s$ values, regardless of whether the true covariance is known or not.

\section{Discussion}\label{section:discussion}

Based on the results of this work - the widths of the SBI posteriors and their coverage fractions measured over many repeated experiments - the good news is that SBI functions as well as covariance estimation methods possibly can. The bad news is the number of simulations required to obtain errors close to the true posterior variance exceeds the computational budget of existing simulation suites, even for data of modest dimension from Gaussian linear models in which the true expectation $\boldsymbol{\xi}[\boldsymbol{\pi}]$ is known. The coverages and posterior widths for SBI presented in this work show that SBI - for modest $n_{\boldsymbol{\xi}}$ and $n_{\boldsymbol{\pi}}$ - does not obtain smaller widths than a Gaussian likelihood analysis with access to an accurate model $\boldsymbol{\xi}[\boldsymbol{\pi}]$.

When considering the posterior widths obtained with SBI - using both NLE and NPE - there is a discrepancy between each compression method. In the limit of a low number of simulations, the linear compression parameterised with the diagonal elements of a simulation-estimated covariance, denoted $\hat{\Sigma}=S_{\text{diag.}}$, is closer to optimal (i.e. a smaller posterior width) in our example than a neural network and the $\hat{\Sigma}=S$ compression - for lower numbers of simulations $n_s$. This is because $S_{\text{diag.}}$ correctly (though limited by $n_s$) estimates the variances of the data but not the covariances - thus reducing the noise propagated by the additonal off-diagonal elements that would be estimated with $S$. This shows that the increase in posterior width obtained by ignoring the off-diagonal  elements of the data covariance (using $\hat{\Sigma}=S_{\text{diag.}}$) is \textit{less} than increase in width when estimating the off-diagonal elements (using $\hat{\Sigma}=S$). 

The posterior variance, when using a neural network for the compression, should be between the $\hat{\Sigma}=S$ and $\hat{\Sigma}=S_{\text{diag.}}$ since the assumption that all of the datavector components are independent is very strong. Ultimately, this depends on the covariance structure of the statistic at hand. How optimal each of the compressions that we test is with respect to each other is not fixed in order. The effects of one particularly unfavourable covariance structure are examined in Appendix \ref{appendix:bad_covariance}. The fact that the posterior variances derived with neural network summaries follow - though are biased above - the Fisher variances (corrected for an estimated covariance via the DS13 factor) stems from the fact that the neural network does not invert any covariance, therefore it can't be biased by minimising the incorrect likelihood in the same way as the $\hat{\Sigma}=S_{\text{diag.}}$ compression. This will depend on the covariance structure of the data. Analysing a DS13-like effect for neural networks remains an objective for future work. None the less, the mean-squared error loss, when minimised in training, does not guarantee an unbiased estimator with the correct variance. It is not clear exactly how the network compression affects the resulting posterior width when optimised with stochastic gradient descent (for low $n_s$) and this problem would not be detected in an analysis that is not directly comparable to an analytic posterior - in particular for the cases in which SBI is needed most.

Our results, based on commonly adopted compression methods, show a concerning inflation of the posterior width compared to the true posterior. In particular for low $n_s$ - values which are comparable to existing simulation budgets available to current analyses - where the widths are greater than those inflated by accounting for an unknown data covariance via the DS13 factor. One exception is in the case the compression only estimates the diagonal elements of true data covariance. It is possible that current analyses based on SBI methods fall in a regime of $n_s$, $n_{\boldsymbol{\xi}}$ and $n_{\boldsymbol{\pi}}$ in which posteriors may only be derived with SBI via compression using a neural network, since the data covariance is singular for $n_s < n_{\boldsymbol{\xi}}$. This shrouds the amount of information that is lost - because the comparison with a linear compression cannot be made - but does not change the fact that for $n_s - n_{\boldsymbol{\xi}} < n_{\boldsymbol{\xi}} - n_{\boldsymbol{\pi}}$ any compression at low $n_s$ is severely sub-optimal. Whilst this is true for our linear model and Gaussian errors, the problem will likely be worse for non-Gaussian errors and non-linear models - i.e. for applications that \textit{require} SBI. As is shown in our posterior widths results (Section \ref{section:results_widths}), the neural network can progress the information return relative to the posterior obtained with a simulation-estimated covariance at a given low value of $n_s$. Despite this improvement, the posterior errors are inflated by a factor of between two and four relative to the true posterior - which increases significantly as a function of $n_{\boldsymbol{\xi}}$ (see Equation \ref{eq:f_ds}) under the assumption that the covariance structure is fixed for increasing $n_{\boldsymbol{\xi}}$. However, if the data covariance structure stays fixed with increasing $n_\xi$, the optimality of the neural network summary may be constant where the DS13 factor would increase - thus decreasing the information in linear summaries using a simulation-estimated covariance compared to those from a neural network. 

The structure of the data covariance significantly affects the information content in summaries from a linear compression, with an estimated covariance, compared to those from a neural network. In Appendix \ref{appendix:bad_covariance} we display results of NPE experiments using an MAF density estimator repeated with a data covariance $\Sigma_r$ that has large off-diagonal elements (unlike the covariance $\Sigma$ used in the other sections of this work). For a covariance with strong correlations between elements in the datavector, a neural network compression is far less optimal - due to the ignorance of the network to the errors on the data - compared to a linear compression with an estimated data covariance $\hat{\Sigma}_r=S$. The same holds for a linear compression parameterised $\hat{\Sigma}_r=S_{\text{diag.}}$ (which minimises the correct $\chi^2$, using the correct model $\xi[\pi]$, but given the wrong data covariance). This highlights the fact that the issue of covariance matrix estimation is not alleviated by data compression. For some statistics with covariances that are unfavourable in this way, the covariance estimation effects on the posterior widths are significant. Using a neural network to compress the data - so as to avoid the estimation of the covariance - does not only not reduce the posterior width relative to the DS13 factor, but it actually substantially deteriorates the returned parameter constraints. 

In the posterior widths of Figure \ref{fig:widths_nle_maf} (see also Appendix \ref{appendix:extra_results}) there is a shift toward lower variance for the reported widths. It should be noted that, in common with other machine learning approaches, the posterior density estimators in the NPE experiments absorb the prior defined by the training set. This requires us to force the same prior for the NLE experiments to obtain a direct comparison between the two approaches. That said, the NLE posteriors (for all compression methods) show a bias to slightly lower posterior widths which can be seen in comparing the $\hat{\Sigma}=\Sigma$ points to the Fisher variance (in Figures \ref{fig:widths_nle_maf} and Appendix \ref{appendix:extra_results}). For the NLE estimators the likelihood function is additionally biased low in posterior width. This is because, similarly to the NPE estimator, the data likelihood is also informed by the prior from which the simulations used to fit the normalizing flows are drawn. It should also be noted that some of the posteriors, for NPE and NLE, are truncated near the prior edges. This occurs more for lower $n_s$ since the estimated covariance causes additional scatter of the MAP - around which the contours are drawn - toward the prior edges.

Compared to analytic methods for either deriving covariance estimators or posteriors that account for a noisy covariance, SBI returns posteriors with correct coverage but larger errors. The posterior errors tend to the Fisher errors at around $4\times10^4$ simulations - for compression and likelihood fitting - meaning that the SBI methods can correctly recover the true posterior given a noisy datavector, though this will depend on the size of the datavector. Comparing the results to the PME estimator \citep{PME}, we see that SBI density estimation requires many more simulations to obtain a similar error width and coverage. For a DES-like datavector \cite{PME} find 400 $N$-body simulations are sufficient to achieve negligible additional statistical uncertainties on parameter constraints from a noisy covariance estimate. Our results show that SBI appears to require many more simulations than the PME estimator. This number will be far greater for LSST and other next-generation surveys - which will also increase further in the presence of nuisance parameters.

Whilst interpreting our results, we note that the DS13 correction term inflates the posterior covariance for a Gaussian likelihood and linear model for the expectation whereas in density estimation SBI approaches, the contour is drawn by a generative model. This estimator for the likelihood shape may suffer from an additional DS13-like term such that the contours drawn in the SBI analyses are inflated on average with respect to the Dodelson-Schneider factor for half the number of simulations ($2n_s$ simulations are required for compression and fitting the flow when $\Sigma$ is unknown). In addition, the scatter of the posterior mean with respect to the true parmaeters may be biased above or below the DS13 scatter of the MAP around the truth. The unknown form of the flow likelihoods may contribute an additional scatter of the posterior mean $\langle\boldsymbol{\pi}\rangle_{\boldsymbol{\pi}|\boldsymbol{\xi}}$ to the true parameters $\boldsymbol{\pi}$. In Appendix \ref{appendix:scatter_mean_map} we measure the scatter of the SBI posterior mean around the true parameters. We fix the true parameters at values in the centre of our prior to minimise effects from the prior which would affect the results of the SBI posterior widths and coverages. There appears to be some minimal scatter for $n_s < 2\times10^3$ which includes simulations to estimate $\hat{\Sigma}$ as well as train the posterior or likelihood estimator.

\section{Conclusions}\label{section:conclusions}

The main result of our paper is illustrated by the combinations of measured posterior coverages with SBI methods, shown in Figure \ref{fig:coverage_nle_maf}, and measured posterior widths, shown in Figure \ref{fig:widths_nle_maf}, for density estimation SBI methods. Appendix \ref{appendix:extra_results} shows further results from the other experiments listed in Table \ref{table:experiments}.

These plots show that the SBI methods do not obtain the expected posterior widths from the analytical prescription of \cite{Dodelson2013} and \cite{Percival2021} when double the number of simulations are input to the experiment. Whilst the methods easily obtain the correct posterior errors $\sigma^2[\pi]$ and coverages at the $1-$ and $2-\sigma$ levels when the covariance is known, the posterior errors are significantly higher than the expected Dodelson-Schneider corrected errors (Equation \ref{eq:dodelson}) for a Gaussian posterior when the data covariance is estimated from an additional set of simulations \citep{Percival2014}. This effect is worst for low numbers of simulations - less than $2 \times 10^3$ in total - and it will only be worsened with the deluge of new higher-dimensional survey data. This implies that the application of current SBI methods to complex non-Gaussian and non-linear statistics (or their combinations together) with additional nuisance parameters may be premature for realistic analyses. Furthermore, were SBI to be used in place of an analytic likelihood, for low numbers of simulations it would not be expected - based on our results - that the parameter constraints would be stronger.

We note that the number of simulations required to obtain the correct posterior errors and coverage is of the same order as the number of simulations provided by simulation suites dedicated to machine learning analyses such as Quijote \citep{Quijote}, CAMELS \citep{CAMELS} and CosmoGrid \citep{cosmogrid}. The exact requirement depends on the survey volume in which statistic is measured. This number of simulations will need to increase to model the covariance, for a realistic volume, of statistics measured in these simulations in order to keep errors within limits required for next-generation surveys. 

To summarise, we find that
\begin{itemize}
    \item posterior estimators from density-estimation SBI methods do not obtain smaller errors than the analytical solution for a Gaussian data likelihood ansatz with a simulation-estimated covariance when 
    \begin{itemize}
        \item the data are drawn from a Gaussian linear model,
        \item the compression is parameterised by a simulation-estimated data covariance, 
        \item an accurate model for the expectation value is used,
        \item and a number of simulations suitable for the analysis of future survey data is used,
    \end{itemize}
\end{itemize}
regardless, these methods
\begin{itemize}
    \item  obtain the correct expected posterior coverage,
    \item correctly estimate the likelihood and posterior shape in the limit of a large number of simulations,
    \item do not show significant scatter of the posterior mean (aside from the DS13 effect) but there is an inflation of the contours (in addition to the DS13 affect) which is not due to density estimation with SBI, but due to SBI’s need for data compression, which in itself will be noisy - since the $\hat{\Sigma}=\Sigma$ experiments obtain the true posterior widths for all $n_s$,
\end{itemize}
and in particular 
\begin{itemize}
    \item SBI `knows' to expand its posterior contour due to the uncertainty in the estimated data covariance which dilutes the parameter constraints - whilst calibrating the coverages - that are derived with such methods,
    \item the errors from SBI methods are significantly larger than the DS13 inflation of errors (in a standard Gaussian likelihood analysis) for of order a few thousand simulations when estimating $\Sigma$ and simulating a training set.
\end{itemize} 

It should be noted that the results of the analyses depend on being able to optimise the density estimators with another set of $10^4$ independent simulations to obtain the hyperparameters of the architecture and training procedure. The set must be independent from the training and validation sets because these sets are used to fit the models and stop their training to avoid overfitting. The linear compression in our experiments would be less optimal for non-Gaussian statistics, further increasing the simulation budget for analyses to reduce the posterior errors. Also, the posterior widths, estimated with SBI when a neural network is used for the data compression, \textit{can} be smaller than the DS13 posterior widths. This is only for a small number of simulations and in the regime that the errors are already much larger than the Fisher errors.

Despite the promise of SBI to seamlessly model combinations of complementary summary statistics, it is not clear based on current SBI methods and density estimation techniques, that more information upon physical parameters of cosmological models can be reliably extracted with the computational resources available to generate the simulations which are required to fit these models. This an important problem in particular because of the cross-correlations that must be modelled with different probes of large-scale structure. Typically large cross-correlations exists between probes in different tomographic bins for example with lensing peaks \citep{chrispeaks} and between smoothing scales in the matter PDF \citep{cora1pt}.

For data from next-generation surveys and analyses that use combinations of probes it is critical to model their correlated signals and systematics to extract as much information as possible from statistics measured at all scales \citep{Krause2017, beyond2pt, Reeves12x2pt}. SBI methods are a promising set of tools that do obtain correctly calibrated posteriors, given enough simulations, that also scale favourably with compute compared to traditional MCMC methods. Despite this, the posteriors obtained by density estimation SBI methods are significantly wider than the naive expectation - given a noisy compression - when no form for the likelihood is assumed and a low number of simulations are used to estimate the data covariance. This inflation of the posterior errors is significant given the typical simulation budgets available for analyses at present. This problem is not alleviated given the optimal linear compression: it remains (and is shifted to either side of the analysis) if the data covariance is not known. Furthermore, depending on the covariance structure, the compression may be insurmountable for neural network based compressions. 

\begin{acknowledgements}
We would like to thank Cora Uhlemann, Kai Lehmann, Sven Krippendorf, and Sankarshana Srinivasan for useful discussions. This work was supported by funding from the Deutsche Forschungsgemeinschaft (DFG, German Research Foundation) under Germany’s Excellence Strategy – EXC-2094 – 390783311. O.F. was supported
by a Fraunhofer-Schwarzschild-Fellowship at Universitäts Sternwarte München.
\end{acknowledgements}

%
%

\bibliographystyle{aa} 
\bibliography{refs} 

\appendix

\section{Normalizing flows: Continuous and Masked autoregressive flows}\label{appendix:flows}

Here we describe two methods we choose to implement normalizing flows that fit either the likelihood or the posterior from simulations and parameters.

\subsection{Masked autoregressive flows} 

To calculate the log-likelihood of the data given the parameters (Equation \ref{eq:flow}) MAFs \citep{mafs} stack a sequence of bijective transformations built using Masked autoencoders (MADEs; \citealt{mades}). A MADE `encodes' an input to a Gaussian distributed variable. The encoding uses a component-wise affine transformation so that the likelihood of the data factorises into a product of Gaussians for each component of the input. 

The affine transformation parameters (a mean and a variance) for each component of the datavector are calculated autoregressively, modelling the PDF of each factorised component as a Gaussian being conditional on the other components preceeding it. The autoregressive factorisation of the likelihood is expressed as 

\begin{align}\label{eq:conditionals}
    \log p_\phi(\boldsymbol{x}|\boldsymbol{\pi})
    &=\sum_{d=1}^D \log p_\phi(x_d|\boldsymbol{x}_{<d}, \boldsymbol{\pi}) \nonumber \\
    &=\sum_{d=1}^D \Bigl (\log \mathcal{G}[f_\phi(\boldsymbol{x}_{<d},\boldsymbol{\pi})|0,1] + \log|\mathbf{J}_{f_\phi}(\boldsymbol{x}_{<d};\boldsymbol{\pi})|\Bigl ),
\end{align}

where each conditional distribution for each dimension $x_d$ depends \textit{only} on part of the input $\boldsymbol{x}_{<d}$, and not on any other dimensions $d'$ where $d < d'\leq D$. If this was not the case, the conditionals would not satisfy the product rule of probability, and a log-likelihood estimate would not be possible. 

The MADE network $f_\phi(\boldsymbol{x};\boldsymbol{\pi})$ ensures that the conditionals are correctly satisfied by masking inputs through the hidden layers. This ensures the output nodes, that parameterise the bijection, are only dependent on the input nodes dictated by the autoregressive factorisation. Since the autoregressive property above is implemented via a product of individual Gaussians, the Jacobian is given by the derivative of each of the output nodes with respect to the input data; 

\begin{align}
    \mathbf{J}_{f_\phi}(\boldsymbol{x};\boldsymbol{\pi})_{ij} &= \frac{\partial f_{\phi, i}(\boldsymbol{x}_{< i}; \boldsymbol{\pi})}{\partial x_j}, \nonumber \\
    &= \frac{\partial}{\partial{x_j}} \frac{x_i - \mu_{\phi,i}(\boldsymbol{x}_{< i};\boldsymbol{\pi})}{\sigma_{\phi,i}(\boldsymbol{x}_{< i};\boldsymbol{\pi})}. 
\end{align}

This shows that the Jacobian matrix is a triangular matrix since the derivative is only non-zero for $i=j$ and $j<i$ terms. This implies that the Jacobian is triangular, meaning that the determinant is simply the product of the diagonal entries, which for the MADE is given by $|\mathbf{J}_{f_\phi}|=\prod_{i=1}^{D}\sigma_{\phi,i}^{-1}(\boldsymbol{x}_{< i};\boldsymbol{\pi})$. The sequence of MADEs used to build the MAF each have random ordering in the autoregressive factorisations of the data likelihood. The determinant of the Jacobian for this sequence of MADEs in the MAF is given by $|\mathbf{J}_{f_\phi}|=\prod_{k=1}^{K}\prod_{i=1}^{D}[\sigma^k_{\phi,i}(\boldsymbol{x}_{< i};\boldsymbol{\pi})]^{-1}$ where $K$ is the number of individual MADE networks.

\subsection{Continuous normalizing flows}

Parameterising a flow without complex modelling of the Jacobian \citep{nvpflows, splineflows, glow} can be done using neural ordinary differential equations \citep{neuralode, diffrax} which model the output of an infinitely deep neural network as the solution to an ordinary differential equation (ODE). As a normalizing flow this ODE maps a sample $\boldsymbol{x}$ from the unknown data distribution to a sample $\boldsymbol{y}$ from a multivariate Gaussian distribution. The path from $\boldsymbol{x}$ to $\boldsymbol{y}$ is parameterised by a `time' variable $t$.

The Neural-ODE and its input is written as 

    \begin{align}
        \dot{\boldsymbol{y}}(t) &= \boldsymbol{f}_\phi(\boldsymbol{y}(t);\boldsymbol{\pi},t) , \nonumber\\
        \boldsymbol{y}(0) &= \boldsymbol{x}.
    \end{align}

We may solve an initial-value problem with this first-order ODE to obtain the state at a later time $T$, denoted  $\boldsymbol{y}(T)$. With an initial value $\boldsymbol{y}(0)=\boldsymbol{x}$, this is written as 

    \begin{align}
        \boldsymbol{y}(T) &= \boldsymbol{y}(0) + \int_{0}^{T} \text{d}s \; \boldsymbol{f}_\phi(\boldsymbol{y}(s); \boldsymbol{\pi}, s) \nonumber \\
        &= \text{ODESolve}(\boldsymbol{y}(0), \boldsymbol{\pi}, \boldsymbol{f}_\phi, 0, T),
    \end{align}

and so a numerical solver can estimate the forward pass of the infinitely deep network which maps a data point $\boldsymbol{x}$ to a latent $\boldsymbol{y}$ given model parameters $\boldsymbol{\pi}$. Here we denote a differential equation solver algorithm as `$\text{ODESolve}$' which in practise is a standard numerical solver. 

The change in log-density from the base distribution to the unknown log-likelihood is calculated by solving another differential equation, known as the \textit{instantaneous change of variables} \citep{neuralode, ffjord}; 

    \begin{equation}
        \frac{\partial}{\partial t} \log p(\boldsymbol{y}(t)|\boldsymbol{\pi}) = -\nabla_{\boldsymbol{y}}\cdot \boldsymbol{f}_\phi(\boldsymbol{y}(t);\boldsymbol{\pi},t),
    \end{equation}

which can be seen as an instance of the Fokker-Planck equation for the time evolution of a random process, where in this case the diffusion coefficient is zero (known as the Liouville equation; \citep{neuralode, diffrax}. This gives the change in log-density when integrated as

    \begin{align}
        \log p(\boldsymbol{y}(T)|\boldsymbol{\pi}) &= \log p(\boldsymbol{y}(0)|\boldsymbol{\pi}) - \int_{0}^{T} \text{d}s \; \nabla_{\boldsymbol{y}}\cdot \boldsymbol{f}_\phi(\boldsymbol{y}(s); \boldsymbol{\pi}, s), \nonumber \\
        &= \log \mathcal{G}[\boldsymbol{y}(0)|\boldsymbol{0}, \mathbb{I}] - \int_{0}^{T} \text{d}s \; \nabla_{\boldsymbol{y}}\cdot \boldsymbol{f}_\phi(\boldsymbol{y}(s); \boldsymbol{\pi}, s), \nonumber \\
        &= \text{ODESolve}(\boldsymbol{y}(0), \boldsymbol{\pi}, \boldsymbol{f}_\phi, 0, T).
    \end{align}

To calculate the likelihood of a sample $\boldsymbol{x}$ given physics parameters $\boldsymbol{\pi}$, an initial value problem is solved. The solution is 

    \begin{equation}\label{eq:cnflogprob}
       \begin{bmatrix}
            \boldsymbol{y}(T) \\
            \log p_\phi(\boldsymbol{x}|\boldsymbol{\pi}) - \log \mathcal{G}(\boldsymbol{y}(T)|\boldsymbol{0}, \mathbb{I}) \\
        \end{bmatrix} 
        =         
        \int_{0}^{T} ds
        \begin{bmatrix}
            \boldsymbol{f}_\phi(\boldsymbol{y}(s); \boldsymbol{\pi}, s) \\
            -\nabla_{\boldsymbol{y}}\cdot\boldsymbol{f}_\phi(\boldsymbol{y}(s); \boldsymbol{\pi}, s)\\
        \end{bmatrix},
    \end{equation}
given the initial values
    \begin{equation}\label{eq:ivp}
       \begin{bmatrix}
            \boldsymbol{y}(0) \\
            \log p_\phi(\boldsymbol{x}|\boldsymbol{\pi}) - \log \mathcal{G}(\boldsymbol{y}(T)|\boldsymbol{0}, \mathbb{I}) \\
        \end{bmatrix} 
        =         
       \begin{bmatrix}
            \boldsymbol{x} \\
            0 \\
        \end{bmatrix}.
    \end{equation}

Which qualitatively amounts to mapping a datapoint to a Gaussian sample whilst calculating the Jacobian along the path. The model is trained via the same maximum-likelihood training described in Section \ref{section:methods} by obtaining the ODE solutions that maximise the log-probability of the data given the model parameters following Equations \ref{eq:ivp} and \ref{eq:cnflogprob}. The solutions can be differentiated due to the implementation of the solver in \texttt{jax} \citep{jax, diffrax}.

\section{Obtaining the best architecture}\label{appendix:hyperparameters}
 We choose the best architecture by sampling a wide variety of architectures and training hyperparameters (batch size, early-stopping patience and learning rate) for individual fits to the likelihood or posterior using MAFs or CNFs. We select the best models by repeating the same experiment with a training set of $10^4$ simulations and an independent test set of $10^4$ simulations to estimate the log-likelihood of the flow. 

The hyperparameters of a likelihood or posterior fit with a CNF in a given experiment conists of parameters for the architecture of the flow model and the optimisation procedure. The parameters for the CNF architecture are

\begin{itemize}
    \item the number of hidden units in the network layers $H\in[8, 16, 32, 64]$,
    \item and the ODE solver timestep width $\text{d}t\in[0.1, 0.05, 0.01]$.
\end{itemize}
and for the MAF architecture they are
\begin{itemize}
    \item the number of layers in the flow network $L\in[1, 2, 3]$,
    \item the number of hidden units in the network layers $H\in[8, 16, 32, 64]$,
    \item and the number of transforms (each of which use a network) $K\in[1, 2, 3, 4, 5]$.
\end{itemize}

\noindent The parameters for the training optimisation are

\begin{itemize}
    \item the number of simulations and parameters in each batch $B\in[40, 50, 60, 70, 80, 90, 100]$,
    \item the learning rate of the optimiser $\eta\in[10^{-5}, ..., 10^{-3}]$ (logarithmic spacing),
    \item and the `patience', the number of epochs required without a decreasing loss before manually stopping the optimisation, $p\in[10, 20, 30, 40, 50]$.
\end{itemize}

We tested two optimisation functions over the repeated experiments. The first was the average flow log-likelihood on a separate test set of $10^4$ simulation and parameter pairs. The architecture and training parameters that maximised this average log-likelihood were chosen and this is the standard method in density estimation SBI analyses in cosmology. The second function was a direct calculation of the KL-divergence between the true analytic Gaussian posterior and the reconstructed posteriors over a set of $J=100$ independently sampled datavectors with $I=2000$ posterior samples each. The estimated KL divergence is expressed as 

\begin{equation}
    \Lambda(\phi) =  \frac{1}{I} \sum_{i=1}^I \frac{1}{J} \sum_{j=1}^J \log \frac{q(\boldsymbol{\pi}_j|\boldsymbol{\hat{x}}_i)}{p_\phi(\boldsymbol{\pi}_j|\boldsymbol{\hat{x}}_i)}.
\end{equation}

Unfortunately this loss calculation is only feasible for the NPE experiments as the posterior samples for NLE require the use of an MCMC sampler on the flow log-likelihood and parameter prior. In practice, we found that for our experiments either optimisation objectives resulted in similar architectures. Optimisation with cross-validations of the validation loss approach was also tested, where independent training and test sets are used to obtain the validation loss. The validation losses from each validation are averaged and this is used as the optimisation function for the hyperparameters. This made no significant difference to the results.

\section{Results of additional experiments}\label{appendix:extra_results}

Here we show the results of the remaining experiments (noted in Table \ref{table:experiments}) that are described in Section \ref{section:results} but not plotted in the main text. In Figures \ref{fig:widths}, we show the marginal variances of the SBI posteriors against the number of simulations $n_s$, for the experiments of NLE with a CNF, NPE with a CNF and NPE with a MAF flow. In Figure \ref{fig:coverage_plots} we show the coverages of the posteriors from the same experiments. Figures \ref{fig:npe_posteriors} and \ref{fig:nle_posteriors} show sets of posteriors derived with varying numbers of simulations $n_s$ for the NPE and NLE methods respectively, using either flow model type. The contours between the NPE, NLE are consistent for each value of $n_s$ as well as with each posterior corrected with the DS13 factor for the value of $n_s$ for each posterior. The MAPs for these experiments, for each panel, scatter considerably for $n_s=1000$ and $n_s=4000$ when an estimated covariance is used. 

    \begin{figure}
    \centering
       \includegraphics[width=8cm]{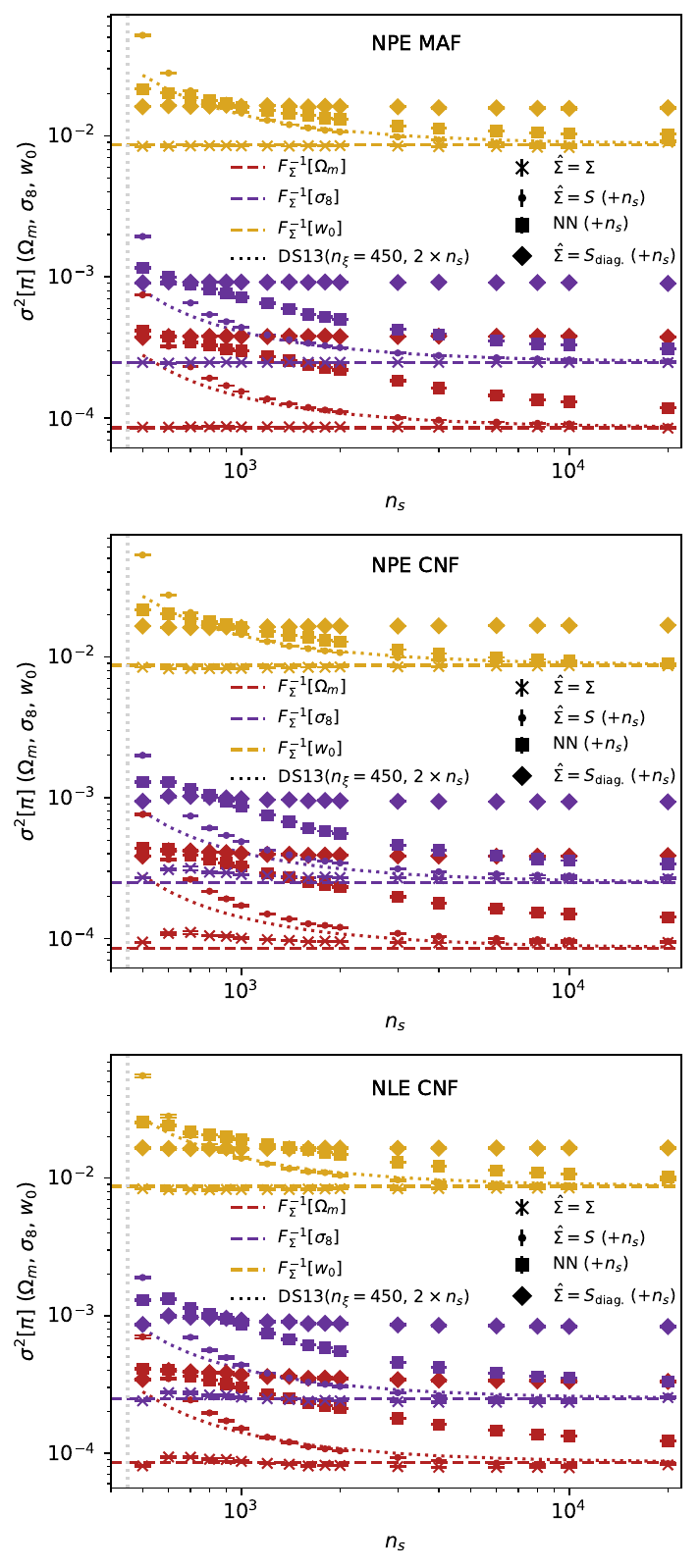}
      \caption{Similar to Figure \ref{fig:coverage_nle_maf}, showing model parameter posterior variance, conditioned on noisy datavectors, estimated with the SBI (the method and normalizing flow model used in each experiment is listed in each panel). The variances between NLE and NPE for each compression are similar except for the small underestimation of the variance in the NLE experiments, due to the likelihood being additionally informed by the prior. The additional simulations, not labeled on the $x$-axis, but required for the separate compressions (where the true covariance is unknown) are noted for each method. When the true data covariance is not known, requiring the use of double the number of simulations, the reconstructed posterior errors from SBI are significantly higher than the \cite{Dodelson2013} corrected errors for less than $4\times10^3$ simulations.}
       \label{fig:widths}
    \end{figure}

    \begin{figure*}
    \centering
       \includegraphics[width=16cm]{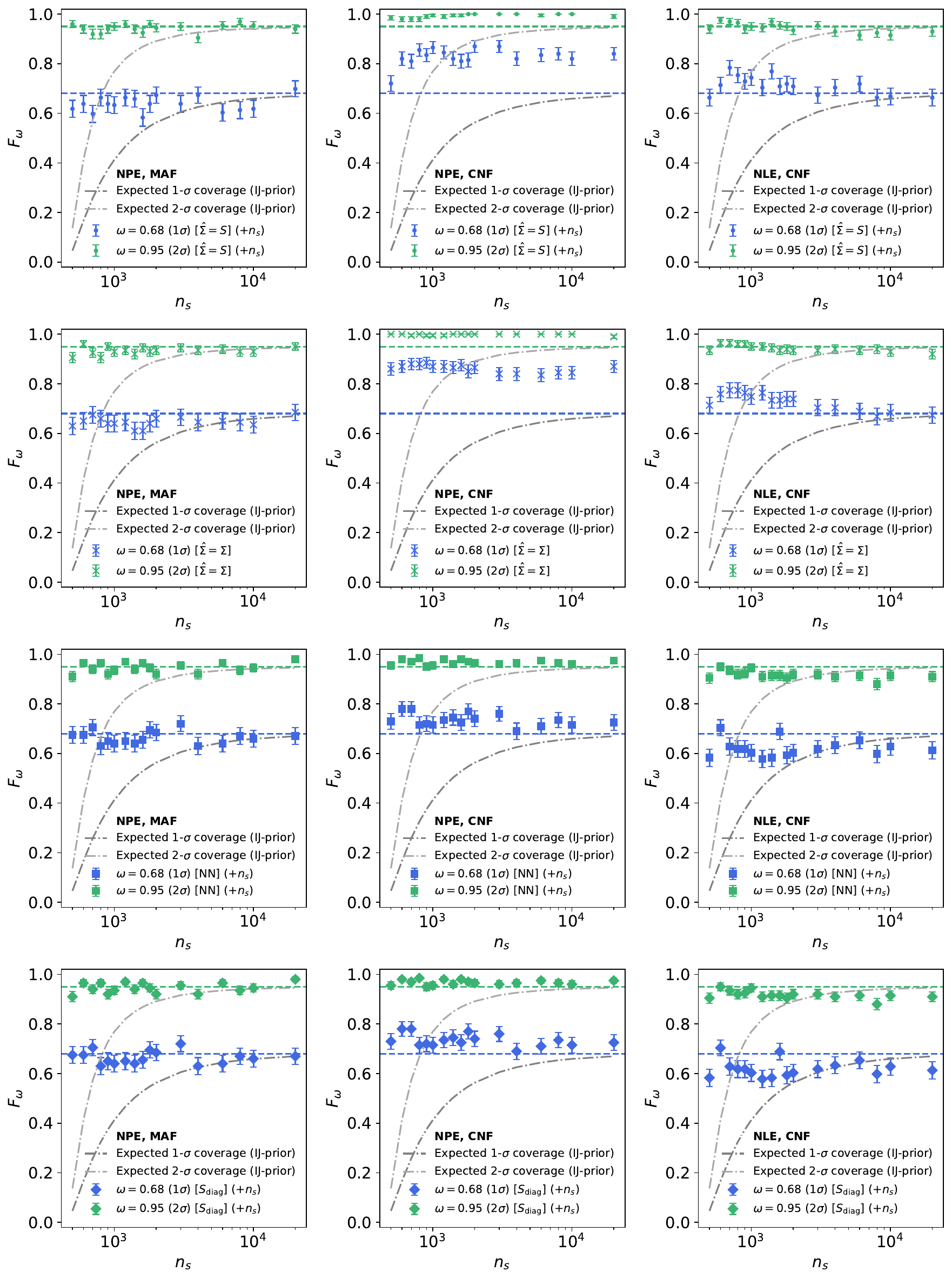}
       \caption{Coverage $F_\omega$, i.e. how often the true cosmology in the experiment is found inside the 68\% ($1-\sigma$) and 95\% ($2-\sigma)$) credible regions of the estimated posterior (see Equation~ \ref{eq:expected_coverage}) against the number of simulations $n_s$ used for the training set. Shown here are the results for Neural Posterior Estimation (separately with a MAF and CNF model) and Neural Likelihood Estimation (with a CNF model) for independently sampled data vectors and data covariance matrices in a series of repeated experiments with the number of simulations for each experiment on the horizontal axis. The expected coverage of a Gaussian posterior with a debiased estimate of the precision matrix (using the Hartlap correction, Equation \ref{eq:HartlapFactor}) and posterior covariance corrected with the \cite{Dodelson2013} factor is plotted for both coverage intervals with dashed lines. The grey-toned lines show the expected coverage of the common approach using a Gaussian posterior with a precision matrix corrected by applying the Hartlap factor. The additional simulations, not labeled on the $x$-axis, but required for the separate compressions (where the true covariance is unknown) are noted for each method. The SBI posteriors obtain the correct coverage to within errors for all numbers of simulations $n_s$ and each compression method.}     
       \label{fig:coverage_plots}
    \end{figure*}

    \begin{figure*}
    \centering
    \includegraphics[width=17cm]{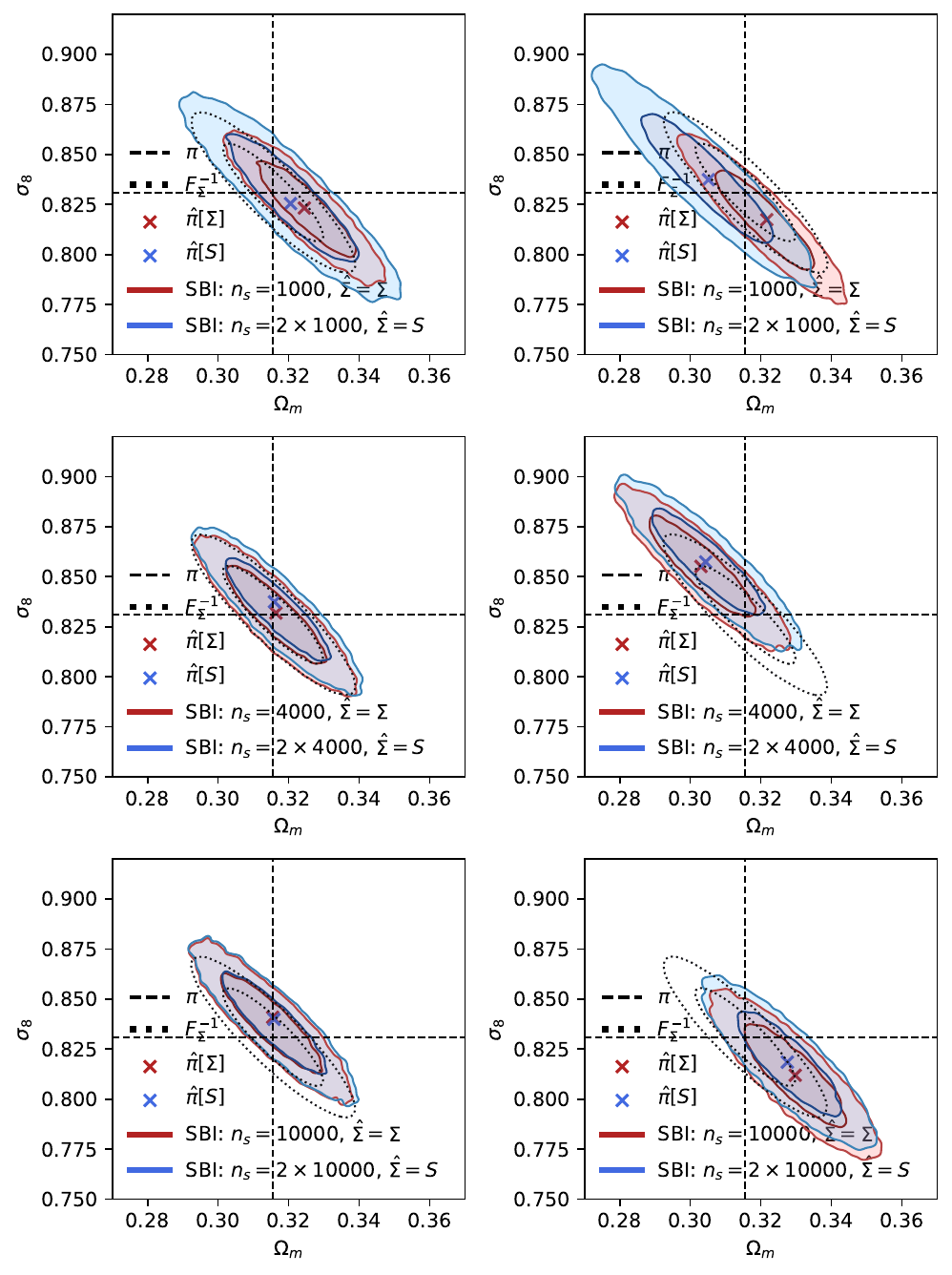}
    \caption{SBI posteriors (red for true covariance $\hat{\Sigma}=\Sigma$, blue for simulation-estimated covariance $\hat{\Sigma}=S$, used in the compression of Equation \ref{eq:freq_map}) derived with Neural Posterior fits from a set of repeated experiments for each value of the number of simulations $n_s$ using a CNF model (left) and a MAF model (right). Each panel is for a different random realization of data $\boldsymbol{\hat{\xi}}$ and covariance $S$, drawn from Gaussian and Wishart distributions, respectively. Datavectors $\boldsymbol{\hat{\xi}}$ linearly compressed to summaries $\boldsymbol{\hat{\pi}}$ are shown in red and blue. A Fisher forecast at true parameters $\pi$ with the true data covariance $\Sigma$ is shown with a dotted black line.}
    \label{fig:npe_posteriors}
    \end{figure*}
    
    \begin{figure*}
    \centering
    \includegraphics[width=17cm]{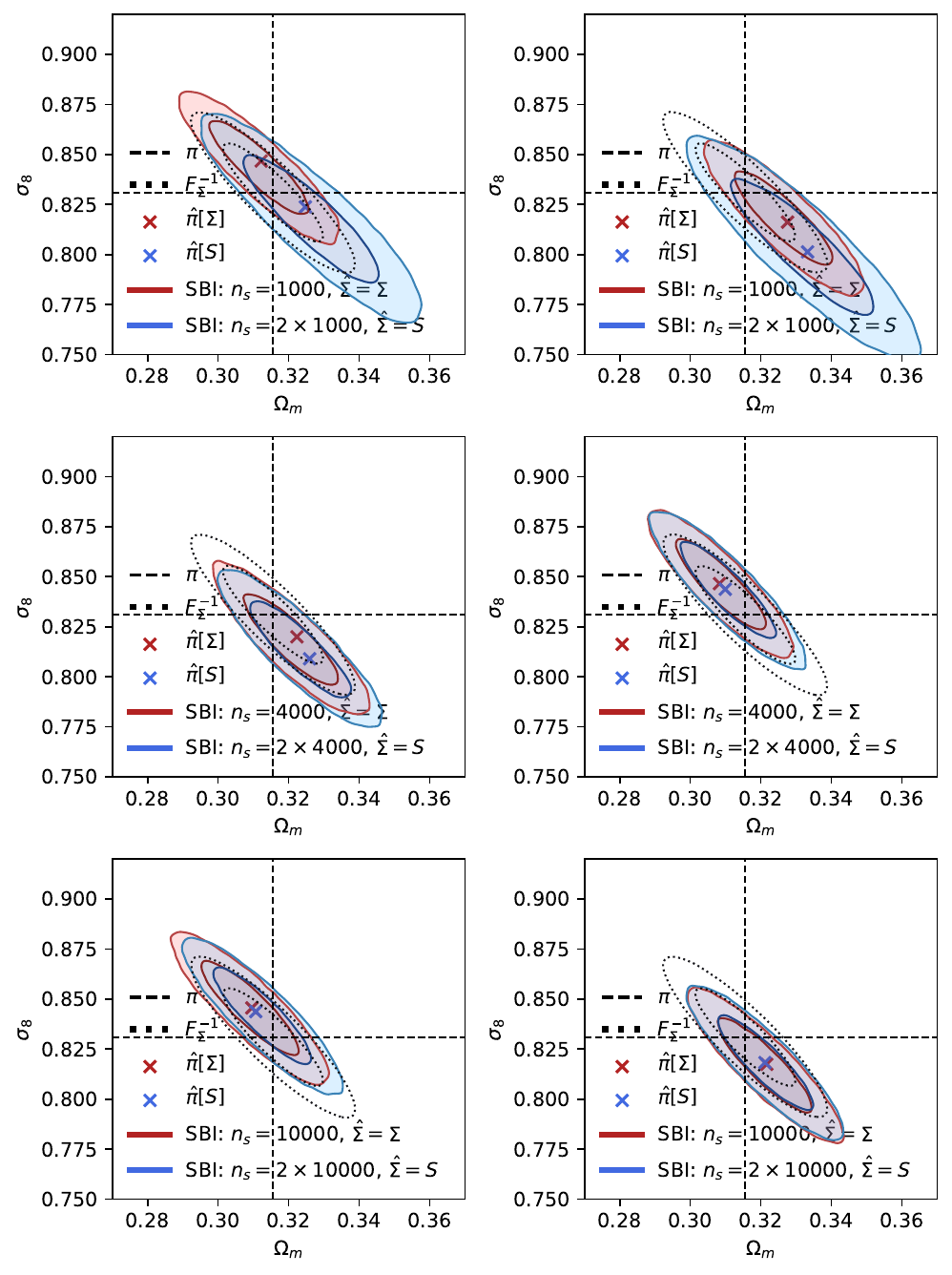}
    \caption{Plot of SBI posteriors (red for $\hat{\Sigma}=\Sigma$, blue for $\hat{\Sigma}=S$, used in the compression of Equation \ref{eq:freq_map}) derived with Neural Likelihood fits from a set of repeated experiments for each value of the number of simulations $n_s$ using a CNF (left) and a MAF (right) model. Each panel is for a different random set of data $\boldsymbol{\hat{\xi}}$ and covariance $S$ drawn from Gaussian and Wishart distributions respectively. Datavectors $\boldsymbol{\hat{\xi}}$ linearly compressed to summaries $\boldsymbol{\hat{x}}$ are shown in red and blue. A Fisher forecast at true parameters $\pi$ with the true data covariance $\Sigma$ is shown in black.}
    \label{fig:nle_posteriors}
    \end{figure*}

\section{Fitting the expectation versus the covariance}\label{appendix:fitting_expectation}
The results of \citet{Dodelson2013} and \cite{Percival2021} depend on the assumption of a linear model $\boldsymbol{\xi}[\boldsymbol{\pi}]$ with Gaussian errors - in particular when the expectation value is known exactly. For the SBI methods in this work that estimate the likelihood or posterior; the expectation, covariance and likelihood shape are all fit from simulations. To isolate the contribution to the posterior width from not knowing the expectation, we fit a model to a noisy datavector where the parameters and the expectation are not known in a Gaussian likelihood analysis. We estimate the mean from a set of simulations and use it to parameterise a Gaussian likelihood with the true covariance matrix (and the same prior used throughout this work). We sample the posterior with a MCMC sampler. The results are shown in Figure \ref{fig:fit_expectation}. This shows that the error contribution is noticeable but far less than that from the noise in the covariance estimate for the same number of simulations. Since the sample mean and covariance are independent \citep{Anderson2003} this test is sufficient to show that the error contribution to the SBI posterior - due to an unknown expectation - does not contribute significantly to the posterior width. This is true for all values of $n_s$ we consider in this work. The increase in width compared to when the covariance is known is due to the unknown data covariance.

\begin{figure}
\centering
\includegraphics[width=0.95\columnwidth]{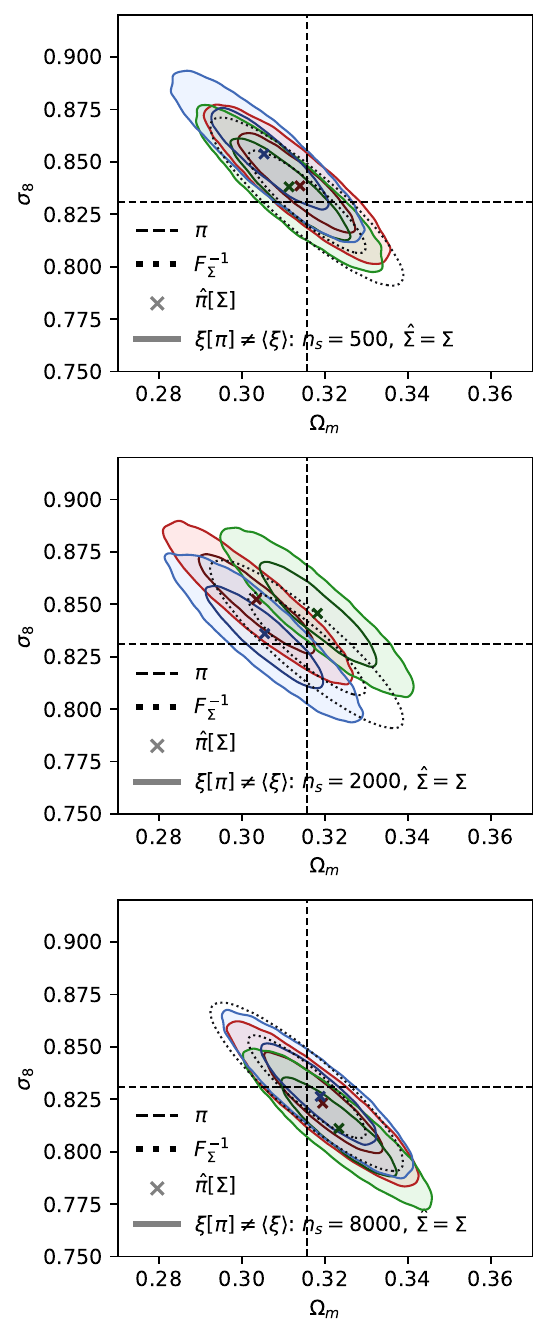}
\caption{Plot of posteriors derived from traditional likelihood analyses using a Gaussian likelihood with the true data covariance matrix, where the expectation $\boldsymbol{\xi}[\boldsymbol{\pi}]$ is estimated from $n_s$ simulations. The posterior samples are obtained from MCMC sampling the analytic posterior. The same prior used in the experiments for this work where the model for the expectation is fit to data alongside the parameters and the data covariance is known. This shows that the error contribution to the posterior from an unknown model is much less than that due to the data covariance being estimated from the same number of simulations.}
\label{fig:fit_expectation}
\end{figure} 

\section{Scatter of posterior means against true parameters}\label{appendix:scatter_mean_map}

In Figure \ref{fig:scatter_mean_map} we show, for all the experiments described in the main text, the scatter of the SBI-posterior means against the true parameters. This is important to measure because whilst the target posterior is a multivariate Gaussian, the SBI posterior is by no means a multivariate Gaussian itself. This means that the posterior mean and the MAP do not necessarily coincide at the same point in parameter space. We measure the squared-difference of the posterior mean against the true parameters as a function of the number of simulations $n_s$ input into the experiment over all the experiments run at each $n_s$ value. The scatter $(\langle \boldsymbol{\pi} \rangle - \boldsymbol{\pi})^2$, if greater than the posterior variance dictated by DS13, suggests that density estimation methods suffer from an additional effect that is not explained simply by the DS13 scattering.

    \begin{figure*}
    \centering
       \includegraphics[width=18cm]{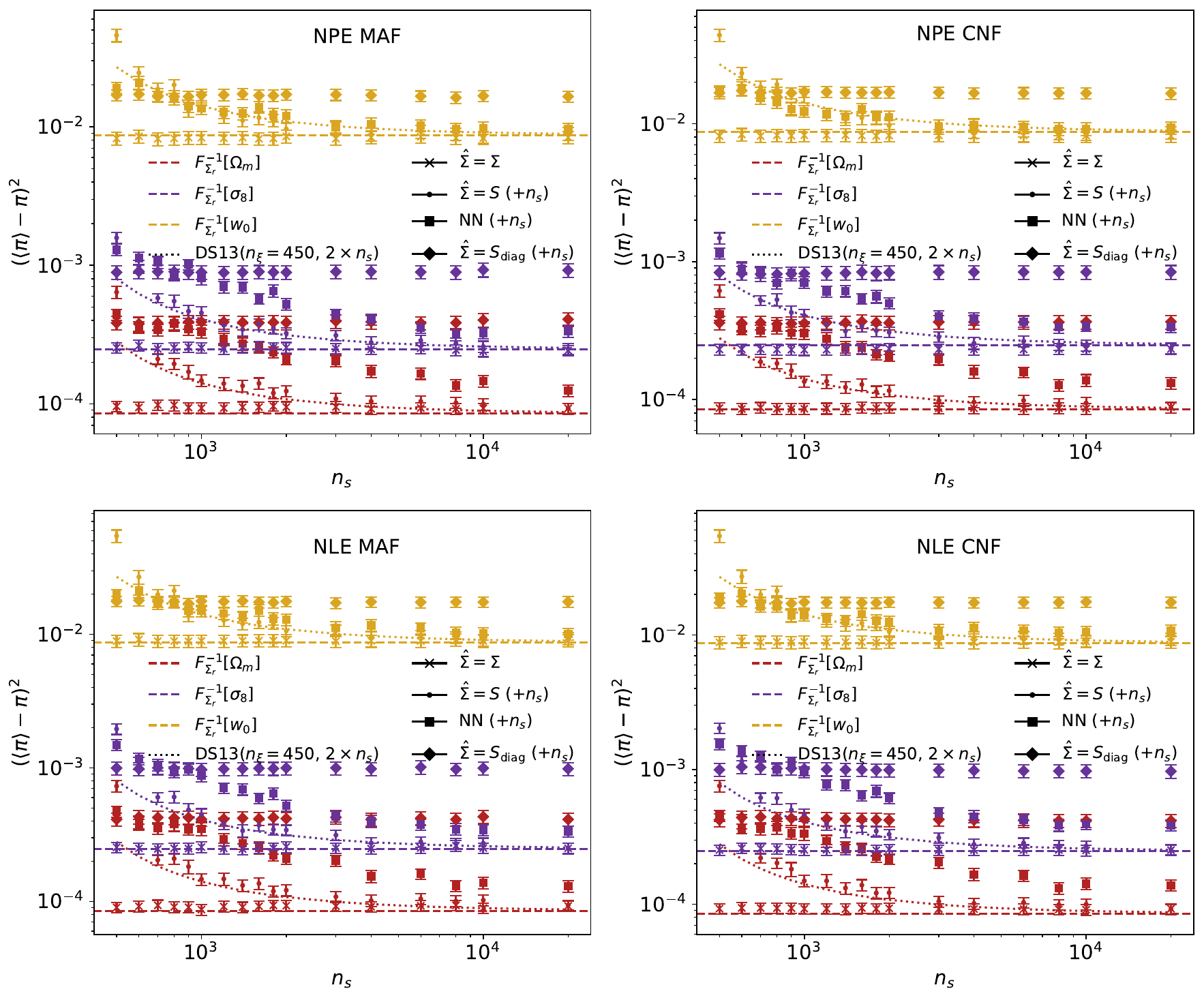}
       \caption{Plot of the scatter in the posterior mean of SBI with respect to the true parameters. The true parameters are fixed to the fiducial parameters $\boldsymbol{\pi}^0$ at approximately the centre of our prior to reduce truncation of the posterior at the prior boundaries which is more significant at lower $n_s$ (see Equation \ref{eq:f_ds}). The bias in the $S_{\text{diag.}}$ estimated covariance is due to ignoring the cross correlations in the datavector - this applies similarly to the neural network (labeled `NN') variances. The similarity of the points in these plots to the plots of the marginal variances of the SBI posteriors against the number of simulations shows that the posterior mean does not scatter significantly more than the MAP.}
       \label{fig:scatter_mean_map}
    \end{figure*}

\section{Experiments with an undesirable covariance matrix}\label{appendix:bad_covariance}

As discussed in Section \ref{section:results}, the structure of the data covariance $\Sigma$ has a significant effect on the reconstructed posterior widths. The effect can be seen in the posterior widths (shown in Figure \ref{fig:coverage_nle_maf} and Appendix \ref{appendix:extra_results}) for SBI posterior estimators (either NLE or NPE) fit to summaries from a linear compression using $\hat{\Sigma}=S_{\text{diag.}}$ or a neural network. This is a distinct effect from the DS13 inflation of posterior contours - which depends on the number of simulations $n_s$ at fixed $n_{\boldsymbol{\xi}}$ - which is caused by the noise in an estimated data covariance. If the covariance structure is such that there are large off-diagonal elements the benefit of a compression with either of these methods vanishes. This is because both methods ignore large cross-correlation elements in the true covariance.}

\begin{figure}
    \centering
       \includegraphics[width=\columnwidth]{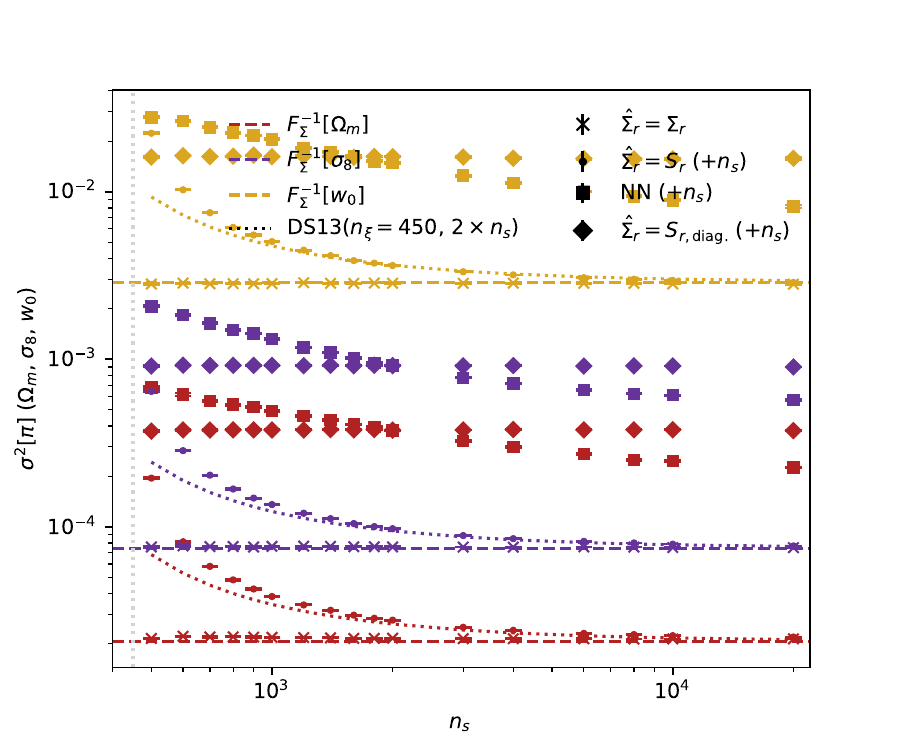}
       
       \caption{Average model parameter posterior variance conditioned on noisy datavectors estimated with Neural Posterior Estimation using a masked autoregressive flow. The same marginal Fisher variances, DS13 factors and compression methods are used for this plot as in Figure \ref{fig:widths_nle_maf}. These results depend on using $\Sigma_r$ as the true covariance (See Section \ref{section:discussion}) for the data generating process. When the true data covariance is not known and has significant non-diagonal elements ($r=0.2$, See Appendix \ref{appendix:bad_covariance}), the compression using either a neural network $f_\psi$ or an estimate of the diagonal elements  $S_{\text{diag.}}$ of the covariance in a linear compression fails catastrophically.}
       \label{fig:widths_bad_cov}
    \end{figure}

To illustrate the effects of such a covariance on the compression methods (and therefore the posteriors derived with SBI) we create a covariance matrix $\Sigma_r$ with large cross correlations between neighbouring elements in the datavector. The matrix has elements

\begin{equation}
    (\Sigma_r)_{ij} = 
    \begin{cases} 
        \Sigma_{ij}, & \text{if } i = j\\
        r\sqrt{\Sigma_{ii}\Sigma_{jj}}, & \text{if } i = j - 1 \text{ or} j = i - 1\\
        0, & \text{if  else} \\
    \end{cases} 
\end{equation}

where $r\in[0, 1)$ is a coefficient that controls the correlation of the datavector components (when this covariance is used to generate data as in our experiments, i.e. as the `true' covariance). As can be seen in Figure \ref{fig:widths_bad_cov} the nature of this covariance means that the benefit of using a neural network (or simply estimating the diagonal elements $S_{\text{diag.}}$ of the covariance for a linear compression) vanishes with the posterior widths being far larger than the DS13 corrected variances for a sampled covariance matrix $\hat{\Sigma}=S$ that estimates all of the elements. The bias is not present in the posterior widths from SBI when the full covariance is estimated for a linear compression - the DS13 inflation of errors persists.

\end{document}